\documentclass[11pt]{article}

\newcommand{\nc}{\newcommand}

\nc{\eqr}[1]{(\ref{#1})}
\nc{\sref}[1]{\S~\ref{#1}}
\nc{\tref}[1]{Table~\ref{#1}}
\nc{\fref}[1]{Figure~\ref{#1}}
\nc{\cref}[1]{Chapter~\ref{#1}}
\nc{\beq}{\begin{equation}}
\nc{\eeq}{\end{equation}}
\nc{\barray}{\begin{eqnarray}}
\nc{\earray}{\end{eqnarray}}
\nc{\barrayn}{\begin{eqnarray*}}
\nc{\earrayn}{\end{eqnarray*}}
\nc{\bcenter}{\begin{center}}
\nc{\ecenter}{\end{center}}

\nc{\Htwo}{H^2(X,\real)}
\nc{\Htwoz}{H^2(X,\Z)}
\nc{\HttpiZ}{H^2(X,\tpi \Z )}
\nc{\Hp}{H^{2}_{\tp}}
\nc{\Hm}{H^{2}_{\tm}}
\nc{\imt}{\mbox{Im }\tau}
\nc{\imtpar}{(\mbox{Im }\tau )}
\nc{\onehalf}{\frac{1}{2}}
\nc{\onefour}{\frac{1}{4}}
\nc{\lra}{\longrightarrow}
\nc{\ra}{\rightarrow}
\nc{\mod}{\mbox{mod }}
\nc{\pf}{{\sc Proof: }}
\nc{\qbar}{\overline{q}}
\nc{\tbar}{\overline{\tau}}
\nc{\slz}{SL(2,\Z)}
\nc{\sltau}{\frac{a\tau +b}{c\tau +d}}
\nc{\tpi}{2\pi}
\nc{\tp}{\mbox{\tiny $+$}}
\nc{\tm}{\mbox{\tiny $-$}}
\nc{\qed}{\mbox{\large$\Box$}}
\nc{\vol}{\mbox{Vol }}
\nc{\ztwo}{\Z_{\mbox{\tiny $2$}}}
\nc{\zth}{\Z_{\mbox{\tiny $3$}}}
\nc{\dm}{\partial_{\mu}}
\nc{\dn}{\partial_{\nu}}
\nc{\dum}{\partial^{\mu}}
\nc{\dun}{\partial^{\nu}}
\nc{\Dm}{D_{\mu}}
\nc{\Dn}{D_{\nu}}
\nc{\Dum}{D^{\mu}}
\nc{\sqrg}{\sqrt{G}}
\nc{\Div}{\nabla\cdot}
\nc{\Curl}{\nabla\times}
\nc{\del}{\partial}
\nc{\tr}{\mbox{tr}}

\nc{\setall}{\setcounter{equation}{0}
        \setcounter{definition}{0}
        \setcounter{lemma}{0}
        \setcounter{convention}{0}
        \setcounter{conjecture}{0}
        \setcounter{theorem}{0}
        \setcounter{assertion}{0}
        \setcounter{property}{0}
        \setcounter{fact}{0}
        \setcounter{corollary}{0}
		\setcounter{claim}{0}}
\nc{\setequation}{\setcounter{equation}{0}}

\nc{\obyt}[2]{\left( \hspace{-1mm}\begin{array}{c}
        #1 \\ #2
        \end{array}\hspace{-1mm}\right)}
\def\sla#1{\raise.15ex\hbox{/}\kern-.57em #1}
\def\slas#1{\raise.15ex\hbox{/}\kern-.62em #1}
\nc{\tbyt}[4]{\left( \begin{array}{rr}
        #1 & #2 \\
        #3 & #4
        \end{array}\right)}
\nc{\abcd}{\left( \begin{array}{cc}
        a & b \\
        c & d
        \end{array}\right)}
\nc{\inner}[2]{\langle #1 , #2 \rangle}
\nc{\e}[1]{{\mbox e}^{#1}}
\nc{\met}[2]{g_{#1 #2}}
\nc{\oover}[1]{\frac{1}{#1}}
\nc{\wed}[2]{ #1 \wedge #2}
\nc{\bhat}[1]{\hat{\mbox{\boldmath $#1$}}}
\nc{\mbold}[1]{\mbox{\boldmath $#1$}}

\newif\ifscrf\scrftrue
\ifx\footscrfont\nullfont
  \scrffalse
\fi
\newif\iffn\fnfalse

\ifscrf
  \font\footscrfont=rsfs10

\fi

\ifscrf
  \def\Scr#1{\iffn
    \mathchoice{\hbox{\footscrfont #1}}{\hbox{\footscrfont #1}}
    {\hbox{\smallfootscrfont #1}}{\hbox{\tinyfootscrfont #1}}\else
    \mathchoice{\hbox{\scrfont #1}}{\hbox{\scrfont #1}}
    {\hbox{\smallscrfont #1}}{\hbox{\tinyscrfont #1}}\fi}
\else
  \def\Scr{\cal}
\fi

\def\SB{{\Scr B}}

\def\SF{{\Scr F}}

\def\SR{{\Scr R}}

\def\ST{{\Scr T}}
\def\SU{{\Scr U}}

\def\sCC{{\kern 0.27em\vrule height1.45ex width0.03em depth0em
	  \kern-0.30em\rm C}}
\def\C{{\mathchoice
  {\sCC}
  {\sCC}
  {\kern 0.225em \vrule height1.05ex width0.025em depth0em \kern-0.25em \rm C}
  {\kern 0.180em \vrule height0.78ex width0.02em depth0em \kern-0.2em \rm C}
	}}
\def\sHH{{\rm I\kern-.16em{}H}}
\def\H{{\mathchoice
  {\sHH}
  {\sHH}
  {\rm I\kern-.13em{}H}
  {\rm I\kern-.13em{}H} }}
\def\sNN{{\rm I\kern-.16em{}N}}
\def\N{{\mathchoice
  {\sNN}
  {\sNN}
  {\rm I\kern-.12em{}N}
  {\rm I\kern-.10em{}N} }}
\def\sPP{{\rm I\kern-.16em{}P}}
\def\P{{\mathchoice
  {\sPP}
  {\sPP}
  {\rm I\kern-.12em{}P}
  {\rm I\kern-.10em{}P} }}
\def\sQQ{{\kern 0.27em \vrule height1.45ex width0.03em depth0em
	  \kern-0.30em \rm Q}}
\def\Q{{\mathchoice
	{\sQQ}
	{\sQQ}
  {\kern 0.225em \vrule height1.05ex width0.025em depth0em \kern-0.25em \rm Q}
  {\kern 0.180em \vrule height0.78ex width0.020em depth0em \kern-0.20em \rm Q}
	}}
\def\sRR{{\rm I\kern-0.16em{}R}}
\def\R{{\mathchoice
  {\sRR}
  {\sRR}
  {\rm I\kern-0.12em{}R}
  {\rm I\kern-0.10em{}R} }}
\def\sZZ{{\rm Z\kern-0.32em{}Z}}
\def\Z{{\mathchoice
  {\sZZ}
  {\sZZ} 
  {\rm Z\kern-0.3em{}Z}     %.3
  {\rm Z\kern-0.25em{}Z} }}  %.25
\def\ZZZ{{\rm Z\kern-0.24em{}Z}}

\let\useblackboard=\iftrue
\newfam\black
\useblackboard
\font\blackboard=msbm10 scaled \magstep1
\font\blackboards=msbm7
\font\blackboardss=msbm5
\textfont\black=\blackboard
\scriptfont\black=\blackboards
\scriptscriptfont\black=\blackboardss
\def\Bbb#1{{\fam\black\relax#1}}
\else
\def\Bbb{\bf}
\fi

\def\yboxit#1#2{\vbox{\hrule height #1 \hbox{\vrule width #1
\vbox{#2}\vrule width #1 }\hrule height #1 }}
\def\fillbox#1{\hbox to #1{\vbox to #1{\vfil}\hfil}}
\def\ybox{{\lower 1.3pt \yboxit{0.4pt}{\fillbox{8pt}}\hskip-0.2pt}}

\def\QZ{\Bbb{Z}}

\nc{\vE}{\vec{E}}
\nc{\vB}{\vec{B}}

\nc{\cA}{{\cal A}}
\nc{\cF}{{\cal F}}
\nc{\cL}{{\cal L}}
\nc{\cN}{{\cal N}}
\nc{\cH}{{\cal H}}
\nc{\cO}{{\cal O}}
\nc{\cR}{{\cal R}}
\nc{\cg}{{\cal G}}
\nc{\cM}{{\cal M}}

\nc{\bA}{{\bf A}}
\nc{\bB}{{\bf B}}
\nc{\bE}{{\bf E}}
\nc{\bI}{{\bf I}}
\nc{\bJ}{{\bf J}}
\nc{\bK}{{\bf K}}
\nc{\bR}{{\bf R}}
\nc{\bZ}{{\bf Z}}

\nc{\al}{\alpha}
\nc{\be}{\beta}
\nc{\ga}{\gamma}
\nc{\de}{\delta}
\nc{\ep}{\epsilon}
\nc{\n}{\nu}
\nc{\m}{\mu}

\nc{\sskip}{\vspace{5mm}}
\nc{\nskip}{\vspace{-2mm}}
\nc{\vs}[1]{\vspace{#1}}
\nc{\hs}{\hspace{1cm}}
\nc{\hshalf}{\hspace{5mm}}

\ifx\undefined\square
  \def\square{\vrule width.6em height.5em depth.1em\relax}\fi
\def\qed{\ifhmode\unskip\nobreak\fi\quad
  \ifmmode\square\else$\m@th\square$\fi}
\newtheorem{lemma}{\large\bf LEMMA}

\newtheorem{theorem}{\large\bf THEOREM}

\newtheorem{conjecture}{\large\bf CONJECTURE}

\newtheorem{claim}{\large\bf CLAIM}

\renewcommand{\thefootnote}{\fnsymbol{footnote}}

\begin{document}
\begin{titlepage}
{\flushright{\small DAMTP-2000-104\\
MIT-CTP-3022\\SLAC-PUB-8619\\SU-ITP-00/20\\hep-th/0009129\\}}

\begin{center}
{\LARGE The Hurwitz Enumeration Problem of Branched 
Covers \\ \vspace{2mm} and Hodge Integrals}
\end{center}

\vspace{1mm}
\begin{center}
{\sc Stefano Monni\footnote{E-mail: S.Monni@damtp.cam.ac.uk.}} \\
{\it DAMTP \\ University of Cambridge \\ Wilberforce Road \\
Cambridge CB3 0WA, U.K.}
\\
\vspace{5mm}
{\sc Jun S. Song\footnote{E-mail:
jssong@mit.edu.  Research supported in part
by an NSF Graduate Fellowship and the U.S. Department of Energy 
under cooperative research
agreement $\#$DE-FC02-94ER40818.}}\\ 
{\it Center for Theoretical Physics\\  Massachusetts 
Institute of Technology\\ Cambridge, MA 02139, U.S.A.}
\\
\vspace{4mm}
and
\vspace{4mm}
\\
{\sc Yun S. Song\footnote{E-mail: yss@leland.stanford.edu.  
Research supported in part by an NSF Graduate Fellowship and the
U.S. Department of Energy under contract DE-AC03-76SF00515.}
}\\ {\it Department of Physics \& SLAC \\ Stanford University \\
Stanford, CA 94305, U.S.A.}
\end{center}

\vspace{2mm}

\begin{abstract}
\noindent
We use  algebraic methods to compute the
simple Hurwitz numbers for arbitrary source and target Riemann
surfaces.   For an elliptic curve target, we reproduce the results
previously obtained by string theorists.  Motivated by the 
Gromov-Witten potentials, we find  a general
generating function for the simple Hurwitz numbers in terms of the
representation theory of the symmetric group $S_n$.  We also
find a generating function for Hodge integrals on the moduli
space $\overline{\cM}_{g,2}$ of Riemann surfaces with two 
marked points, similar to that found by Faber and Pandharipande
for the case of one marked point.

\end{abstract}
\end{titlepage}

\renewcommand{\thefootnote}{\arabic{footnote}}
\setcounter{footnote}{0}
%%%%%%%%%%%%
%
%%%%%%%%%%%%
\section{Introduction}
Many classical problems in enumerative geometry have been receiving
renewed interests in recent years,  the main reason being 
that they can be
translated into the modern language of Gromov-Witten theory and,
moreover,  that they can be consequently solved.
One such classical problem which has been under recent active
 investigation is the
Hurwitz enumeration problem of counting topologically distinct,
almost simple, ramified covers of the projective line or more generally
of any Riemann surface. The almost simplicity
condition is that the branch points be  all simple with the possible
exception of one degenerate point, often called $\infty$, the
branching type of whose pre-images being
specified by an ordered partition $\alpha$ of the degree $n$ of the
covering.
Let $\alpha = (\alpha_1, \ldots, \alpha_w)$ be an ordered partition of
$n$, denoted by $\alpha\vdash n$, of length $|\alpha| =w$.  Then,
 the number $r$ of simple branch points is determined by the
Riemann-Hurwitz formula to be:

\beq\label{eq:RH}
        r = (1-2h) n + w + 2g -2 \ ,
        \eeq

\noindent
where  $h$ and $g$ are
the genera of the target and the source Riemann surfaces,
respectively. The number  $\mu^{g,n}_{h,w}(\alpha)$  of
such covers is called the almost simple Hurwitz number, and
in this paper, we mostly restrict ourselves to simple
Hurwitz numbers  $\mu^{g,n}_{h,n}(1^n)$, for which there is no
ramification over $\infty$.
Hurwitz numbers appear in many branches of mathematics and
physics.  In particular, 
they arise naturally in combinatorics, as they count
factorizations of permutations into transpositions, and the original
idea of Hurwitz expresses them in terms of the
representation theory of the symmetric group.  Indeed in this respect,
the most
general problem of counting  covers of Riemann surfaces by Riemann
surfaces, both reducible and irreducible, with arbitrary branch types, 
 has  been completely solved by Mednykh
\cite{Mednykh1, Mednykh}.  His formulas
however generally do not allow explicit computations of the numbers, except
in a few cases. 

It turns out that one can successfully  obtain the
simple Hurwitz numbers using Mednykh's works,
 and in this paper, we shall compute them
at low degrees for arbitrary target and source Riemann surfaces.
 Hurwitz numbers also appear in
physics: when the target is an elliptic curve, they are\footnote{Up to
over-all normalization constants.} the
coefficients in the expansion of the free energies of the large $N$
two-dimensional quantum 
Yang-Mills theory on the elliptic curve, which has in fact a string
theory interpretation \cite{gross}.  The total free
energy and the partition function, which is its exponential,  
can be thought of as  
generating
functions for simple Hurwitz numbers $\mu^{g,n}_{1,n}$.  
Generalizing this analogy, we have determined the
generating functions for arbitrary targets in terms of the 
representation theory of the symmetric group $S_n$. 

In the framework of
Gromov-Witten theory, simple Hurwitz numbers  
can be considered as certain cohomological classes
evaluated over the virtual fundamental class of the moduli space of
stable maps to $\P^1$ \cite{FanP}.  By exploiting
this reformulation, many  new results
such as new recursion relations \cite{FanP, JSS}  have  been  obtained.
  Furthermore,   a beautiful
link with Hodge integrals has been discovered, both by virtual
localization \cite {FanP,  GrV} and  by other methods
\cite{ELSV}.  It is therefore natural to expect that the knowledge of
Hurwitz numbers might  be used to gain new insights  into
Hodge integrals.  This line of investigations has previously
led to a closed-form
formula for a generating function for Hodge integrals over the moduli
space $\overline{\cM}_{g,1}$ of 
curves with one marked point \cite{ELSV,FP}.  Similarly,  
in this paper, we consider the following generating function for
Hodge integrals over $\overline{\cM}_{g,2}$:
        \beq
	 G (t,k) := {1\over 2}+ \sum_{g\geq 1}\, t^{2g}\,
\sum_{i=0}^{g} \, k^i  \int_{\overline{{\cal M}}_{g,2}} 
		\frac{\lambda_{g-i}}{(1-\psi_1)(1-\psi_2)}\, .
	\eeq
For negative integral
values of $k$, we have  managed  to compute $G(t,k)$ in a
closed form by relating the integrals to the almost simple Hurwitz
numbers $\mu^{g,2k}_{0,2}(k,k)$.  We then conjecture a simplified
version of our rigorously obtained result, and this conjectural
counterpart can then be analytically continued to all values of $k$.
We have checked that the conjectural form of our formula holds true
for $-60\leq k \leq 1$, but unfortunately, we have not been able to
prove it for arbitrary $k.$
The success of
the computation makes us speculate that in more general cases, similar
results  might be within reach, and the simplicity of the results
suggests  that new yet undiscovered structures might be present.

This paper is organized as follows:  
in \S\ref{sec:comp}, we briefly explain
the work of Mednykh and apply it to compute the
simple Hurwitz numbers; in
\S\ref{sec:gen}, we find  the generating functions for 
all simple Hurwitz numbers; 
\S\ref{sec:hodge} discusses our closed-form formula for the
generating function for Hodge integrals over $\overline{\cM}_{g,2}$; 
and, we conclude by drawing the reader's attention to some important
open questions.

%%%%%%%%%%%%%%%%%%%%%%
\newpage
\noindent
{\bf NOTATIONS:}  We here summarize our notations to be used
throughout the paper:

\vspace{5mm}
\begin{tabular}{lll}
$\mu^{g,n}_{h,n}$ & \hspace{5mm}&  \parbox[t]{4in}{The usual
 degree-$n$ simple Hurwitz
numbers for covers of a genus-$h$ Riemann surface by genus-$g$ Riemann
surfaces. }\vspace{4mm} \\
$\tilde{\mu}^{g,n}_{h,n},\, N_{n,h,r}$ &\hspace{5mm}& \parbox[t]{4in}{Mednykh's
definition of 
simple Hurwitz numbers, including the fixed point contributions
of the $S_n$ action. (See \S\ref{subsec:Mednykh} for details.)
}\vspace{4mm} \\
${\SR_n}$ &\hspace{5mm}& \parbox[t]{4in}{The set of all ordinary 
irreducible
representations of the 
symmetric group $S_n$.}\vspace{4mm} \\ 
$\chi_{\gamma} (1^{\alpha_1}\cdots n^{\alpha_n})$ &\hspace{5mm}&
\parbox[t]{4in}{The
character of the irreducible representation $\gamma\in {\SR_n}$ evaluated
at the conjugacy class $[(1^{\alpha_1}\cdots n^{\alpha_n})]$.  For
those $\alpha_i$ which are zero, we omit the associated cycle in our
notation.}\vspace{4mm} \\ 
$f^{\gamma}$ &\hspace{5mm} & \parbox[t]{4in}{The dimension of the
irreducible representation $\gamma\in {\SR_n}$. } \vspace{4mm} \\ 
$\SB_{n,h,\sigma}$ &\hspace{5mm} & See \eqr{def:B}.\vspace{4mm} \\ 
$\ST_{n,h,\sigma}$ &\hspace{5mm} & \parbox[t]{4in}{Subset of 
$\SB_{n,h,\sigma}$,  generating a transitive subgroup of $S_n$.}
\vspace{4mm} \\
$H^g_h$ &\hspace{5mm} & \parbox[t]{4in}{Generating functions for
$\mu^{g,n}_{h,n}$, for fixed $g$ and $h$. See \eqr{eq:Hgh}.}
\vspace{4mm} \\ 
$\widetilde{H}^g_h$ &\hspace{5mm} & \parbox[t]{4in}{Generating functions for
$\tilde{\mu}^{g,n}_{h,n}$, for fixed $g$ and $h$.}
\vspace{4mm} \\ 
$H_{g,n}$ &\hspace{5mm} & \parbox[t]{4in}{Simplified notation 
for $\mu^{g,n}_{0,n}$.  Not to be confused with $H^g_h$.}
	\vspace{4mm}\\
$t^p_k$ &\hspace{5mm} &  Entries of the branching type matrix $\sigma$.
	\vspace{4mm}\\
$\hat{t}^i_j$ &\hspace{5mm} & \parbox[t]{4in}{Coordinates on
the large phase space in the Gromov-Witten theory.}
\end{tabular}

\vspace{5mm}
\noindent
In this paper, all simple Hurwitz numbers count irreducible covers, 
unless specified otherwise.

\newpage
%%%%%%%%%%%%%%################################
%
%%%%%%%%%%%%%%################################

\setall
\section{Computations of Simple Hurwitz Numbers}\label{sec:comp}
This section describes our computations of 
 the simple Hurwitz numbers $\tilde{\mu}^{g,n}_{h,n}$.  The simple covers of
 an elliptic curve by elliptic curves are actually unramified, and
we obtain the numbers $\tilde{\mu}^{1,n}_{1,n}$ by using the standard
theory of two-dimensional lattices\footnote{We thank R. Vakil for
 explaining this approach to us.}.  For other values of 
$g$ and $h$, we simplify
 the general formulas of Mednykh \cite{Mednykh1} and explicitly
 compute the numbers for low degrees.

\subsection{Unramified Covers of a Torus by Tori}
For covers of an elliptic curve by elliptic curves, the
Riemann-Hurwitz formula \eqr{eq:RH} becomes
	\beq
	r= w-n \, ;
	\eeq
but since $n \geq w$, there cannot be any simple branch points and the
special point $\infty$  also has no branching.    
As a result, the computation for this case reduces to  
determining the number of degree $n$ unbranched covers of an elliptic
curve by elliptic curves.  
Equivalently, for a given lattice $L$ associated with the target 
elliptic curve, we need to find the number of inequivalent sublattices
$L'\subset 
L$ of index $[L : L'] =n$.   The answer is given by
Lemma~\ref{lemma:1} to be 
	\beq	
	{\tilde\mu}^{1,n}_{1,n} = \sigma_{1} (n) \, ,
	\eeq
where, as usual, $\sigma_k (n) = \sum_{d | n} d^k$.  Note that we are doing 
the actual counting of distinct covers, and our answer 
${\tilde\mu}^{1,n}_{1,n}$ is not equal to $\mu^{1,n}_{1,n}$ which is
defined by incorporating the automorphism group
of the cover differently.  This point will become clear in 
our ensuing discussions.

The generating
function for the  number of inequivalent simple covers
of an elliptic curve by  elliptic curves is thus
given by 
	\beq \label{eq:g=1 h=1}
	{\widetilde{H}}^1_1 = 
\sigma_{1} (n) q^n = - \left(\frac{d \log \eta (q)}{dt} -
\frac{1}{24}\right) \, ,  
	\eeq
where $q = e^t$.

Up to the constant $1/24$, our answer \eqr{eq:g=1 h=1} is a
derivative of the genus-1 free energy $\SF_1$ of string theory on
an elliptic curve target space.  
The expression 
\eqr{eq:g=1 h=1} can also be obtained by counting distinct
orbits of the action of $S_n$ on a set $\ST_{n,1,0}$, which will be
discussed subsequently.  The string theory computation of
 $\SF_1$, however, counts the
number $\mu^{1,n}_{1,n} :=| \ST_{n,1,0}|/n!$ 
without taking the fixed points of the
$S_n$ action into account, and it is somewhat surprising that our
counting is related to the string theory answer by simple 
multiplication by the degree.  It turns out that this phenomenon occurs
for $g=1$ because the function $\sigma_1(n)$ can be expressed as a sum 
of products of $\pi (k)$, where $\pi (k)$ is the number of distinct
partitions of the integer $k$ into positive integers, and because
this sum
precisely appears in the definition of $T_{n,1,0} =|\ST_{n,1,0}| $. 
We will elaborate upon this point in \S\ref{subsec:caution}.
 In other cases,
the two numbers $\mu^{g,n}_{h,n}$ and  $\tilde{\mu}^{g,n}_{h,n}$ are
related by an additive term which generally depends on $g,h$, and
$n$.

%%%%%%%%%%%%%################################
%          Low Degree Computations
%%%%%%%%%%%%%%################################
\subsection{Low Degree Computations from the Work of Mednykh}
\label{subsec:Mednykh} 
The most general
Hurwitz enumeration problem for an arbitrary branch type
has been formally solved by Mednykh in \cite{Mednykh1}.   His answers
are based on the original idea of Hurwitz of reformulating the ramified
covers in terms of the representation theory of $S_n$ \cite{Hurwitz}.   Let
$f:\Sigma_g \ra \Sigma_h$ be a degree-$n$ branched cover of
a Riemann surface of genus-$h$ by a Riemann surface
 of genus-$g$, with $r$ branch points, the orders of whose 
pre-images being specified by the
 partitions $\alpha^{(p)} = (1^{t^p_1}, \ldots,
n^{t^p_n})\vdash n,$ $ p=1, \ldots, r.$  The ramification type of the
covering $f$ is then denoted by the matrix $\sigma= (t^p_s)$.  Two
such branched covers $f_1$ and $f_2$ 
 are equivalent if there exists a homeomorphism
$\varphi: \Sigma_g \ra \Sigma_g$ such that $f_2 = f_1 \circ
\varphi$.  Now, define the set $\SB_{n,h,\sigma}$ as
	\barray
	\SB_{n,h,\sigma} &=& \left\{ 
(a_1, b_1, \ldots, a_h,b_h, (1^{t^1_1}, \ldots,
n^{t^1_n}) , \ldots, (1^{t^r_1}, \ldots,
n^{t^r_n}) \in (S_n)^{2h+r}\,\|\right. \nonumber\\
	&& \hspace{2cm}  \, \prod^{h}_{i=1}\,[a_i, b_i]\,
\prod^{r}_{p=1}\left. (1^{t^p_1}, \ldots, n^{t^p_n}) =1 \right\}
\, , \label{def:B}
	\earray
and $\ST_{n,h,\sigma}\subset \SB_{n,h,\sigma}$
 as the subset whose
elements generate transitive subgroups of $S_n$.  Then, according to
Hurwitz, there is a one-to-one correspondence between irreducible
branched covers and elements of $\ST_{n,h,\sigma}$.  Furthermore,
the equivalence relation of covers gets translated into
conjugation by a permutation in $S_n$, i.e. two elements of 
$\ST_{n,h,\sigma}$ are now
equivalent iff they are conjugates.  Thus, the Hurwitz enumeration
problem reduces to counting the number of orbits in $\ST_{n,h,\sigma}$
under the action of $S_n$ by conjugation.

Let us denote the orders of the sets by
$B_{n,h,\sigma}=|\SB_{n,h,\sigma}|$
and $T_{n,h,\sigma}= |\ST_{n,h,\sigma}|$.  Then, using the classical
Burnside's formula, Mednykh obtains the following theorem for the
number $N_{n,h,\sigma}$ of orbits:

	\begin{theorem}[Mednykh] The number of degree-$n$
 	non-equivalent branched covers of the ramification type
$\sigma = (t^p_s),$ for $ p=1,\ldots, r,$ and $ s=1,\ldots, n$ is
given by
	\barray \label{eq:Master}
		N_{n,h,\sigma} &=& \frac{1}{n} \sum_{ \mbox{\tiny
        $\begin{array}{c} \ell | v\\ m\ell =n\end{array}$}}
\sum_{\frac{1}{(t,\ell)} | d | \ell} \frac{\mu ({\ell\over d}) \,
d^{(2h-2+r)m+1}}{(m-1)!} \left[ \,\sum_{j^s_{k,p}}
T_{n,h,(s^p_k)}\right. \times \nonumber\\
	&& \left. \hspace{1cm}\times\, 
	\sum^{d}_{x=1}\, \prod_{s,k,p} \left[ {\Phi
(x,s/k)\over d}\right]^{j^s_{k,p}} \,\prod_{k,p} {s^p_k \choose {
j^1_{k,p} , \ldots, j^{md}_{k,p}}}\right] 
	\earray
where $t := 
GCD \{ t^p_s \}$, $v: = GCD \{ s\, t^p_s\}$, $(t,\ell) = GCD(t,\ell)$,
$s^p_k =  \sum^{md}_{s=1} j^s_{k,p}$, and 
the sum over $j^s_{k,p}$ ranges
over all collections $\{ j^s_{k,p} \}$ satisfying  the condition
	\beq
	\sum_{ \mbox{\tiny
        $\begin{array}{c} 1\leq k \leq st^p_s/\ell \\ (s/(s,d)) | k|s 
	\end{array}$}} k\, j^s_{k,p} = {s\, t^p_s \over\ell}
	\eeq	
where $j^s_{k,p}$ is non-zero only for $ 1\leq k \leq st^p_s/\ell$
and $(s/(s,d)) | k|s$. The functions $\mu$ and $\Phi$ are the
M\"{o}bius and von Sterneck functions.
	\end{theorem}

As is apparent from its daunting form, the expression
 involves many conditional 
sums and 
does not immediately yield  the desired numerical answers.  Mednykh's works, 
even though quite remarkable, are thus of dormant nature 
for obtaining the closed-form numerical 
answers\footnote{Recently, closed-form answers for
coverings of a Riemann sphere by genus-0,1,2 Riemann surfaces with one
non-simple 
branching have been obtained in \cite{GJV}.} of the Hurwitz
enumeration problem.   

Interestingly, the general formula \eqr{eq:Master} still has some 
applicability.  For example,
in \cite{Mednykh}, Mednykh considers the special case of branch points
whose orders  
are all equal to the degree of the cover and obtains a simplified
formula which is  
suitable for practical applications.  In a similar vein, we discover
that for  
simple branched covers, Mednykh's formula simplifies dramatically and that
for some low degrees, we are able to obtain closed-form answers for
simple Hurwitz numbers of ramified 
coverings of genus-$h$ Riemann surfaces by genus-$g$  Riemann surfaces.

\subsubsection{The Simplifications for Simple Hurwitz Numbers}

We consider degree-$n$ simple branch covers of a genus-$h$ Riemann surface by 
genus-$g$ Riemann surfaces.
A simple branch point has order $(1^{n-2},2)$, and thus the branch type is
characterized by the matrix $\sigma =
(t^p_s)$, for $p =1,\ldots, r,$ and $s=1,\ldots, n,$ where
	\beq
		t^p_s = (n-2) \delta_{s,1} + \delta_{s,2}. 
	\eeq
To apply Mednykh's master formula \eqr{eq:Master}, we need to determine $t = 
GCD \{ t^p_s \}$ and $v= GCD \{ s\, t^p_s\}$, which are easily seen to be
	\beq
		t = 1 \hspace{1cm} \mbox{and} \hspace{1cm} v = 
		\left\{ \begin{array}{ll} 
	2 & \mbox{ for $n$  even\, ,} \\ 
		1 & \mbox{ for $n$ odd\, .}
	\end{array}\right.  
	\eeq
Because $v$ determines the range of the first sum in the master formula, 
we need to distinguish when the degree $n$  is odd or even.

\subsubsection{Odd Degree Covers}
For degree-$n$ odd, we have $\ell = d = (t,\ell) =1$ and $m=n$.  
The constraints $(s/(s,d)) | k|s$ and 
	\beq
		\sum_{1\leq k  \leq s \, t^p_s/\ell} k \, j^s_{k,p}
		= \frac{s \, t^p_s}{\ell}
	\eeq
then determine the collection $\{ j^s_{k,p} \}$ to be
	\beq
		j^s_{k,p} = t^p_s \, \delta_{k,s} \, .
	\eeq
Noting that $\Phi (1,1) =1$, we see that 
the master formula now reduces to
	\beq 
		N_{n,h,\sigma} = 
		\frac{T_{n,h,(s^p_k)}}{n!} 
		\hspace{1cm} 
			\mbox{($n$ odd)}\, , 
		\label{eq:n odd}
	\eeq 
where 
	\beq
		s^p_k = \sum^{n}_{s=1} j^s_{k,p} = t^p_k  = 
		(n-2) \, \delta_{k,1} + \delta_{k,2} \, . \label{eq:s}
	\eeq

%%%%%%%%%%%%%%%%%%%
%
%%%%%%%%%%%%%%%%%%%

\subsubsection{Even Degree Covers}
For degree-$n$ even, $v=2$ and thus $\ell = 1$ or $2$.

\vspace{5mm}
\noindent
\underline{$\ell=1$}: \ The  variables take the same values as in the
case of $n$ odd, and the $\ell =1$ contribution to
$N_{n,h,\sigma}$ is thus precisely given by \eqr{eq:n odd}.

\vspace{5mm}
\noindent
\underline{$\ell =2$}: \ In this case, the summed variables are fixed to be
	\beq
	m = \frac{n}{2}  \hspace{1cm} \mbox{and} 
	\hspace{1cm} d = \ell =2\, .
	\eeq
Then, one determines that
	\beq
	j^s_{k,p} = \frac{t^p_1}{2} \delta_{s,1} \delta_{k,1} +
		t^p_2 \delta_{s,2} \delta_{k,1} \ ,
	\eeq
from which it follows that
	\beq
		\tilde{s}^p_k = \frac{n}{2}\, \delta_{k,1} \ ,
	\eeq
where we have put a tilde over $s^p_k$ to distinguish them 
from \eqr{eq:s}.	
Using the fact that the number $r$ of simple branch points is even, and the
values $\Phi (2,1) = \Phi (2,2) = - \Phi (1,2) =1$, one can now
show that the
$\ell =2$  contribution to $N_{n,h,\sigma}$ is 
	\beq
	\frac{2^{(h -1)n +1}}{(\frac{n}{2} -1)!} \, \left( \frac{n}{2}
	\right)^{r-1}  
	T_{\frac{n}{2}, h, (\tilde{s}^p_k)} \ .
	\eeq
The sum of both contributions is finally given by
	\beq \label{eq:Neven}
	N_{n, h, \sigma} = \frac{1}{n!}\, T_{n,h,(s^p_k)} + 
	\frac{2^{(h -1)n +1}}{(\frac{n}{2} -1)!} \, \left( \frac{n}{2}
	\right)^{r-1}  
	T_{\frac{n}{2}, h, (\tilde{s}^p_k)} \hspace{1cm} 
	\mbox{($n$ even)}\, . \label{eq:n even}
	\eeq	

\vspace{5mm}
\noindent
{\bf NOTATIONS:}  For simple branch types, i.e. for
$\sigma = (t^p_k)$ where $t^p_k = (n-2) 
\delta_{k,1} + \delta_{k,2}$, for $p=1,\ldots, r$ and
$k=1,\ldots, n$, we will
use the notation $T_{n,h,\sigma} =: T_{n,h,r}$.
\vspace{5mm}

The computations of fixed-degree-$n$ 
simple Hurwitz numbers are thus reduced 
to computing the two numbers $T_{n,h,(s^p_k)}$ 
and	$T_{\frac{n}{2}, h, (\tilde{s}^p_k)}$, only the former 
being relevant when $n$ is odd.  
We now compute these numbers for some low degrees and arbitrary genera 
$h$ and $g$.  The nature of the computations is such that we only need to
know the characters of the identity and the transposition elements in $S_n$.

The term $T_{\frac{n}{2}, h, (\tilde{s}^p_k)}$ can be easily 
computed:
	\begin{lemma} \label{lemma:Tltwo} Let $\tilde{s}^p_k = n
\delta_{k,1}$.  Then, 
		\beq
			T_{n, h, (\tilde{s}^p_k)} =
			n! \sum^n_{k=1} \frac{(-1)^{k+1}}{k}
\,\sum_{n_1 + \cdots 
			+ n_k = n} \prod^k_{i=1} \left[
			\sum_{\gamma\in \SR_{n_i}} \left(
\frac{n_i!}{f^{\gamma}} 
			\right)^{2h-2}\right]\ .
	 	\eeq
		where $n_i$ are positive integers, $\SR_{n_i}$ the set of all 
		irreducible representations of $S_{n_i}$, and $f^{\gamma}$ 
		the dimension of the representation $\gamma$.
	\end{lemma}

For $h=0$, we can explicitly evaluate this contribution:
\begin{lemma}
Let $\tilde{s}^p_k = n \delta_{k,1}$.  Then,
        \beq \label{eq:zero}
        T_{n, 0, (\tilde{s}^p_k)} = \sum^n_{k=1}
        \frac{(-1)^{k+1}}{k} \sum_{\mbox{\tiny $\begin{array}{c} n_1+\cdots
        +n_k=n
\\ n_i >0\end{array}$}}  \obyt{n}{n_1,\ldots, n_k} = \left\{ \begin{array}{l}
        1, \,\mbox{\em for } n=1 \\  0, \,\mbox{\em for } n>1 \
        . \end{array} \right.
        \eeq
\end{lemma}
\noindent
{\sc Proof:} The first equality follows from the fact that the
order of a finite group is equal to the sum of squares of
the dimension of its irreducible representations.  The second equality
follows by noticing that
 the expression for $T_{n,0, (\tilde{s}^p_k)}/n!$  is the
$n$-th coefficient of the formal
$q$-expansion of $\log ( \sum^{\infty}_{n=0} q^n/n! )$, which is a
fancy way of writing $q$. {\hfill $\qed$}

Using \eqr{eq:n odd} and \eqr{eq:n even} we have computed closed-form
formulas for the simple Hurwitz numbers for arbitrary source and
target Riemann surfaces for degrees less than 8 in Appendix C.

%%%%%%%%%%%%%%%%%%%%
%
%%%%%%%%%%%%%%%%%%%%

\subsection{Cautionary Remarks}\label{subsec:caution}
Hurwitz numbers are sometimes {\it defined}\/ to be 
$T_{n,h,\sigma}/n!$, counting orbits as if there were no fixed points 
of the action $S_n$ on $\ST_{n,h,\sigma}$. 
The master formula obtained 
by Mednykh uses the Burnside's formula to account for the fixed points.
In the case of simple Hurwitz numbers, this will lead to
an apparent discrepancy between our results and those obtained
by others for even degree covers, the precise reason
being that for even degree covers, say of degree-$2n$, the action 
of $(2^{n})\in S_{2n}$  on $\ST_{2n,h,\sigma}$ has fixed points which 
are counted by the second term in \eqr{eq:Neven}.  Consequently, to 
obtain the usual even degree Hurwitz numbers, we just need to consider 
the contribution of the first term in \eqr{eq:Neven}.  For odd degree
cases, there is no non-trivial fixed points, and our formula needs
no adjustment.  The following examples of the discussion would be
instructive:  

%%%%%%%%%%%%%%
\newpage
	\noindent
	\underline{{\sc Example One}}

\noindent
Let us explicitly 
count the double covers of an elliptic curve by
genus-$g$ Riemann surfaces.  The set $\ST_{2,1,2g-2}$
is given by	
	\beq
	\ST_{2,1,2g-2} = \left\{ (a,b,(2)^{2g-2}) \in S^{2g}_{2}\,\|
		\, aba^{-1}b^{-1} (2)^{2g-2} = 1
		\right\}\ .
	\eeq
Since $S_2$ is commutative and $(2)^2 =1$, any pair 
$(a,b)\in S_2\times S_2$ satisfies the required condition.  Hence,
the order of $\ST_{2,1,2g-2}$ is four.  Now, to count
non-equivalent coverings, we need to consider the action of $S_2$ 
on the set $\ST_{2,1,2g-2}$ by conjugation.  Again, since $S_2$
is abelian, it is clear that it acts trivially on the set and thus
that there are 4 inequivalent double covers of an elliptic curve
by genus-$g$ Riemann surfaces.  The commonly adopted definition of
Hurwitz number, however, specifies that we should take the order of the
set $\ST_{2,1,2g-2}$ and divide it by the dimension of $S_2$, 
yielding 2 as its answer.  This number 2 is precisely the
first contribution in the Burnside's formula:
	\beq
		N_{2,1,{2g-2}} = \frac{1}{|S_2|} \sum_{\sigma\in S_2}
		| F_{\sigma}| = \frac{F_{(1^2)} }{2!} + 
		\frac{F_{(2)}}{2!} =2+2 =4
	\eeq
where $|F_{\sigma}|$ is the order of the fixed-point set under 
the action of $\sigma\in S_2$.  For odd $n$, $S_n$ acts freely
on the set $\ST_{n,h,r}$, but for even $n$, it has fixed points
and our formula \eqr{eq:Neven} accounts for the phenomenon, 
truly {\it counting}\/ the number of inequivalent covers.

To avoid possible confusions, we thus use the following 
notations to distinguish the two numbers:
	\beq
	\mu^{g,n}_{h,n} := {T_{n,h,r}\over n!}  , \ \ \ \mbox{ for all $n$},
	\eeq
and
	\beq
		\tilde{\mu}^{g,n}_{h,n} := N_{n,h,r}\, .
	\eeq
It turns out that current researchers are mostly interested in 
$\mu^{g,n}_{h,n}$; for example, it is this definition of simple
Hurwitz numbers that appears in the string theory literature and
in relation to Gromov-Witten invariants.  In this paper, we will
compute the numbers
$\tilde{\mu}^{g,n}_{h,n}$ and indicate the $\ell=2$
contributions which can be subtracted to yield $\mu^{g,n}_{h,n}$.
We will however find generating functions only for the case
$\mu^{g,n}_{h,n}$.

%%%%%%%%%%%

	\vspace{5mm}
	\noindent
	\underline{{\sc Example Two}}

\noindent
The above discussion shows
 that the two numbers $\tilde{\mu}^{g,n}_{h,n}$ and
$\mu^{g,n}_{h,n}$
differ by the second term in \eqr{eq:Neven} and thus
are not related by simple multiplicative factors.  For $h=1$ and $g=1$,
however, we have previously observed that
${\widetilde{H}}^1_1$  given in
\eqr{eq:g=1 h=1} is equal to $\partial_t \SF_1$, up to an additive
constant, implying that
	\beq
	\tilde{\mu}^{1,n}_{1,n} = n\, \mu^{1,n}_{1,n}\ .
	\eeq
This special equality actually
follows from Lemma~\ref{lemma:sigma_1}, since
we know that $\tilde{\mu}^{1,n}_{1,n} = \sigma_1 (n)$ and since one
can show that
	\beq
	\mu^{1,n}_{1,n} := \frac{T_{n,1,0}}{n!} = \sum^n_{k=1} 
	\frac{(-1)^{k+1}}{k} \sum_{m_1 + \cdots + m_k =n} 
	\left(\prod^{k}_{i=1}\pi (m_i) 
			\right)\ .
	\eeq

%%%%%%%%%%%%%%%%############################################
%
%%%%%%%%%%%%%%%%############################################

\subsection{Recursive Solutions to $T_{n,h,r}$ for an Elliptic Curve
($h=1)$} \label{sec: recursion}
Elliptic curve is the simplest Calabi-Yau manifold and is of
particular interest to string theorists.  The free energies
$\SF_g$ count the numbers $\mu^{g,n}_{h,n}$, and string
theorists have computed $\SF_g$ for $g\leq 8$ \cite{Rudd}.
Using the approach described in the previous subsection,
we have obtained the closed-form formulas for
$N_{n,h,r}$ for $n < 8$.  For $h=1$, its $\ell =1$ parts
agree with the known free energies $\SF_g$.
Although our results are rewarding
in that they give explicit answers for all $g$ and $h$,
further computations become somewhat cumbersome beyond degree 8.
For higher degrees, we therefore adopt a recursive method to solve
$T_{n,1,r}$ on a case-by-case basis.
	
The number of reducible covers $B_{n,h,\sigma}/n!$  and that of
irreducible covers $T_{n,h,\sigma}/n!$ are related by exponentiation
\cite{Mednykh}:
        \beq \label{eq:b=et}
                \sum_{\sigma \geq 0} \frac{B_{n,h,\sigma}}{n!} w^{\sigma}
                =\exp \left(
                \sum_{\sigma \geq 0} \frac{T_{n,h,\sigma}}{n!} w^{\sigma}
                \right) \ ,
        \eeq
where $w^{\sigma}$ denotes   the multi-product
        \beq
           w^{\sigma} :=  
		   \prod_{p=1}^{r} \prod_{k=1}^{n} \,w^{t^p_k}_{pk}
        \eeq
in the indeterminates $w_{pk}$ and $\sigma \geq 0$ means $t^p_k \geq 0,
\forall\, p,k$.  From \eqr{eq:b=et}, one can derive
       \beq
                B_{n,h,\sigma} = \sum^n_{k=1} \frac{1}{k!}
                \left(\hspace{-1mm}\begin{array}{c}
                n\\ n_1, \ldots, n_k \end{array}\hspace{-1mm}\right) \,
                \sum_{ \mbox{\tiny $\begin{array}{c} n_1+\cdots +n_k=n \\ \sigma_1 +\cdots +
                \sigma_k =\sigma\end{array}$}} T_{n_1,h,\sigma_1}
                \cdots T_{n_k,h,\sigma_k} \ .
        \eeq
In particular, for simple covers of an elliptic curve, partitioning
$\sigma$ appropriately yields
        \beq \label{eq:F_g recursion}
                \, \frac{T_{n,1,r}}{n!} = \frac{B_{n,1,r}}{n!} -
	\sum^n_{k=2} \frac{1}{k!} 
                                \sum_{ \mbox{\tiny $\begin{array}{c}
	n_1+\cdots +n_k=n \\ 2\ell_1 +\cdots 
	+  2\ell_k =r\end{array}$}}
        \left(\hspace{-1mm}\begin{array}{c}
         r\\ 2\ell_1,\ldots,2\ell_k \end{array}
         \hspace{-1mm}\right)\prod_{i=1}^{k}
         \frac{T_{n_i,1,\mbox{\tiny 2$\ell_i$}}}{n_i!} \ ,
        \eeq
where $n_i$ and $\ell_i$
are positive and non-negative integers, respectively.
For fixed degree $n$, \eqr{eq:F_g recursion} expresses
$T_{n,1,r}$ in terms of lower degree and
lower genera Hurwitz numbers, and $B_{n,1,r}$.  The number $B_{n,1,r}$ in
this case reduces to
        \beq
        B_{n,1,r}=\frac{n!}{2^r} \, \sum^n_{k=1}\sum_{ \mbox{\tiny
        $\begin{array}{c} n_1+\cdots +n_k=n
        \\ n_1 \geq n_2 \geq \cdots \geq n_k\end{array}$}} \left[
        \sum_{i\in {\cal I}} p_i (p_i -1) \prod_{j\neq i}\left(
        \frac{p_i-2 -p_j}{p_i-p_j}\right)
        \right]^r\, ,
        \eeq
where $p_i = n_i +k -i$ and ${\cal I} = \{i\in \{ 1, \ldots, k\}\,
\| \,    (p_i-2) \geq 0\}.$  In Appendix~\ref{app:red},
 we provide the explicit values of $B_{n,h,r}$ for $n \leq 10$.

We have implemented the recursion into a 
{\em Mathematica} program which,
using our results from the previous
subsection as inputs, computes $T_{n,1,r}$ for
$n\geq 8$.  For the sake of demonstration, 
we present some numerical values of
$T_{n,1,r}/n!$  for $n \leq 10$ in
Appendix~\ref{app:elliptic curve}.

%%%%%%%%%%%%%################################################
%    
%%%%%%%%%%%%%################################################
\setall
\section{Generating Functions for Simple Hurwitz Numbers}
\label{sec:gen}

Recently, G\"{o}ttsche has conjectured an expression for the
generating function for the
number of nodal curves on a surface  $S$,
with a very ample line bundle $L$, in terms of certain
universal power series and 
basic invariants \cite{Goettsche}.  More precisely, he conjectures that
the generating function $T(S,L)$ for the number of nodal
curves may have the form
	\beq
	T(S, L) = \exp\,( c_2(S) \,A  +  K_S^2 \,B + K_S\cdot L \,C + 
        L^2 \,D)\, ,
	\eeq
where $A,B,C,D$ are universal power series in some formal variables and
$K_S$ the canonical line bundle of $S$.

In a kindred spirit, it would be interesting to see whether such 
universal structures exist for Hurwitz numbers.  For a curve,
the analogues of $K_S$ and $c_2(S)$ would be the genus of the
target and $L$ the degree of the branched cover.  It turns out
that for simple Hurwitz numbers, we are able to find their
generating functions in closed-forms, but the resulting 
structure is seen to be 
more complicated than that for the case of surfaces.

%%%%%%%%%%%%
%
%%%%%%%%%%%%

\subsection{Summing up the String Coupling Expansions} \label{subsec:sum}
The free energies $\SF_g$ on an elliptic curve have been computed
in \cite{Rudd} up to $g=8$, and their $q$-expansions\footnote{Here,
 $q=\exp (\hat{t})$, where $\hat{t}$ is a formal variable dual to the K\"{a}hler
class of the elliptic curve.} 
agree precisely with our results shown in Appendix \ref{app:elliptic
curve}.   

For a fixed degree $n<8$, we know $\SF_g$ for all $g$, so we can sum
up the expansion
	\beq\label{eq:Ftotal}
	\SF = \sum_g \lambda^{2g-2}\, \SF_g \ ,
	\eeq
up to the given degree $n$ in the world-sheet instanton expansion.
That is, we are summing up the string coupling expansions, and 
this computation is a counterpart of ``summing up the world-sheet
instantons'' which string theorists are accustomed to studying.

Consider the 
following generating function for simple Hurwitz numbers:
	\beq \label{eq:Phi}
	\Phi (h) = \sum_{g,n} \frac{T_{n,h,r}}{n!}\,
 \frac{\lambda^r}{r!}\, q^n=\sum_{g,n} \mu^{g,n}_{h,n}\,
 \frac{\lambda^r}{r!} \, q^n  \ , 
	\eeq
which coincides with the total free energy \eqr{eq:Ftotal} for $h=1$.
For low degree 
simple covers of an elliptic curve, we can use our results \eqr{eq:e1} 
to perform the
summation over the number $r$ of simple branch points and get
	\barray
	\Phi (1) &=& \sum_{g} \, \lambda^{2g-2}\,  \SF_g\nonumber\\
	&=&-\,\log (q^{-1/24} \, \eta (q)) + 2\,\left[ \cosh (\lambda)
	-1\right] q^2  
	+2\, \left[\cosh(3\lambda) -\cosh (\lambda)\right] q^3 \nonumber\\
	&&+\, 2\, \left[ \cosh (6\lambda) + \frac{1}{2} \cosh (2\lambda)
	-\cosh (3\lambda) + \cosh(\lambda) - \frac{3}{2} \right] q^4
\nonumber\\ 
	&&+\, 2\, \left[ 1 + \cosh(10\lambda) -\cosh(6\lambda) + \cosh
(5\lambda) - 
	\cosh(4\lambda) + \cosh (3\lambda) - 2 \, \cosh (\lambda)
	\right]q^5\nonumber\\
	&& +\, 2 \left[ \cosh(15\lambda) -\cosh (10\lambda) + \cosh(9\lambda)
	- \cosh(7\lambda) + \frac{1}{2} \cosh(6 \lambda) -
\cosh(5\lambda) \right. 
	\nonumber\\
	&&+\,\left . 
	2 \cosh(4\lambda) -\frac{2}{3} \cosh(3\lambda) + \frac{1}{2} 
	\cosh(2\lambda) + 2 \cosh(\lambda) - \frac{11}{3}
	\right]q^6 + {\cal O}(q^7) \ .
	\earray
The partition function $Z= e^{\Phi (1)}$ is then given by
	\barray
		Z &=&  1 + q + 2\,\cosh (\lambda)\,{q^2} 
 	+ \left[ 1 + 2\,\cosh (3\lambda) \right] \,{q^3} + 
   \left[ 1 + 2\,\cosh (2\lambda) + 2\,\cosh (6\lambda) \right]
\,{q^4}\nonumber\\ 
	 && +\, 
   \left[ 1 + 2\,\cosh (2\lambda) + 2\,\cosh (5\lambda) + 2\,
\cosh (10\lambda) \right] \,
    {q^5}\nonumber\\
	&&  +\, {{\left[ 1 +4\,\cosh (3\lambda) +2 \cosh 
		(5\lambda)+ 2\,\cosh (9\lambda) + 
         2\,\cosh (15\lambda) \right]}}\,{q^6} \nonumber\\
         &&+\, \left[ 1 + 2 \cosh (\lambda) + 2 \cosh (3\lambda) + 2
\cosh(6\lambda)  
	 + 2 \cosh (7\lambda) + 2 \cosh(9\lambda)
 + 2 \cosh(14\lambda)\right. \nonumber\\	
 	&&\left. +\, 2 \cosh(21 \lambda) \right] q^7 +
	2\, \left[ 1+  \cosh (2 \lambda) + 2 \cosh (4\lambda) +
\cosh(7\lambda)  
	 + \cosh (8\lambda) +  \cosh(10\lambda)\right. \nonumber\\
 &&\left. +\,  \cosh(12\lambda) +  \cosh(14 \lambda) + \cosh (20 \lambda) 
 + \cosh (28 \lambda) \right] q^8
	+{{\cal O}(q^9)}.
	\earray
At this point, we can observe a pattern emerging, and indeed,
the partition function can be 
obtained to all degrees from
the following statement which, we subsequently discovered, 
was also given in \cite{Dijk}:
	\begin{claim}  \label{claim:dannazione} The partition function $Z$, or
	the exponential of the generating function for simple Hurwitz numbers, 
	for an elliptic curve target is given by
	\beq
	Z = 1+ q + \sum_{n\geq 2} \left( \sum_{\gamma\in {\SR_n}} \cosh \left[ 
	\left(\hspace{-1mm}\begin{array}{l}
	n\\2  \end{array}\hspace{-1mm}\right)
	\frac{\chi_{\gamma}(2)}{f^{\gamma}} \,\lambda
	\right]\right) q^n \, .\label{eq:Z (1)}
	\eeq
	\end{claim}
\noindent
{\sc Proof:}  From \eqr{eq:F_g recursion}, we see that
	\beq \label{eq:BFg}
	\frac{B_{n,1,r}}{n!\, r!} = 
	\sum^n_{k=1} \frac{1}{k!} 
             \sum_{ \mbox{\tiny $\begin{array}{c}
	n_1+\cdots +n_k=n \\ 2\ell_1 +\cdots +
                          2\ell_k =r\end{array}$}}\prod_{i=1}^{k}
           (\SF_{\ell_i+1})_{q^{n_i}}\ ,	
	\eeq
where, as before, $n_i$ and $\ell_i$ are positive and non-negative integers,
respectively, and
$(\SF_g)_{q^m}$ is the coefficient of $q^m$ in the genus-$g$ free energy.  The
numbers $B_{n,1,r}$ are determined to be
	\beq
	B_{n,1,r} = \left\{ \begin{array}{ll}
		n!\, \left(\hspace{-1mm}\begin{array}{l}
			n\\2 \end{array}\hspace{-1mm}\right)^r \, \left[	
			\sum_{\gamma\in {\SR_n}}  \left(
			\frac{\,\chi_{\gamma}(2)}{f^{\gamma}}\right)^r
		\right]  & \mbox{, for $n\geq 2$} \\
			\delta_{r,0}  & \mbox{, for $n\leq 1$\, .}
			\end{array} \,  \right.
	\eeq
Now, multiplying both sides of \eqr{eq:BFg} by $\lambda^r q^n$ and summing
over all even $r\geq 0 $ and all $n\geq 0$ proves the claim.
{\hfill  $\qed$}

\noindent
The  argument of hyperbolic-cosine is known as the central character
of the irreducible representation $\gamma$ and can be evaluated as in
\eqr{eq:argument}.

\vspace{.5cm}
%%%%%%%%%%%%%%%%%%%%%%%%%%%%%%%%%%%%%%%%%%%%%%%%%%%%%%%%%%%%%%%%%%%%%
\noindent
\underline{Further Recursions for Closed-Form Answers}\\
The above explicit form of the partition function gives rise to a
powerful way of recursively solving for 
 the simple Hurwitz numbers
$\mu^{g,n}_{1,n}$ for a given degree $n$, similar to those given in
\eqr{eq:e1}.  Let us consider this more
closely.  Suppose that, knowing closed-form formulas for
$\mu^{g_i,n_i}_{1,n_i}$ for all $n_i < n$ and arbitrary $g_i$, we are
interested in deriving a closed-form formula for $\mu^{g,n}_{1,n}$,
where $g$ is again arbitrary.  The key idea is to match the
coefficient of $\lambda^{2g-2} q^n$ in the expansion of the partition
function $Z$ with the coefficient of the same term in the expansion 
\[
\exp[ \Phi(1)] \ = \ 1 + \Phi(1) + {1\over 2}[\Phi(1)]^2 + \cdots + {1\over
k!} [\Phi(1)]^k + \cdots .
\]
The coefficient of $\lambda^{2g-2}q^n$ in $\Phi(1)$ contains precisely
what we are 
looking for, namely $\mu^{g,n}_{1,n}$.  On the other hand, the
coefficients of $\lambda^{2g-2}q^n$ in $[\Phi(1)]^k$, for $k>1$, are
given 
in terms of  
$\mu^{g_i,n_i}_{1,n_i}$, where $n_i < n$ and $g_i \leq g$.
But, by hypothesis, we know $\mu^{g_i,n_i}_{1,n_i}$ for all $n_i < n$,
and therefore we can solve for  $\mu^{g,n}_{1,n}$ in a closed-form.
Using this method, we have obtained the
degree-8  Hurwitz numbers, and the answer agrees with the known
results as well as the
computation done by our earlier recursive method.

This recursive method also works for determining the general simple
Hurwitz numbers $\mu^{g,n}_{h,n}$, upon using the general ``partition
function'' \eqr{eq:Z (h)} in place of $Z$.

%%%%%%%%%%%%%%%%%%
%
%%%%%%%%%%%%%%%%%%

\subsection{The Generating Functions for Target Curves of Arbitrary Genus}
For arbitrary genus targets, there is a natural generalization of the
above discussion on the generating functions.
We have previously defined the generating function $\Phi (h)$ 
to be
	\[
	\Phi (h) = \sum_{r,n\geq 0} \mu^{g,n}_{h,n} 
\frac{\lambda^r}{r!} q^n \ ,
	\]
and seen that for $h=1$, it coincides with the
total free energy of string theory on an elliptic curve target, 
where $\lambda$ is identified with the string coupling constant.
For $h\neq  1$, however, the formal parameter $\lambda$ should be
actually viewed as the parameter\footnote{Unfortunately, we have
previously used the notation $t^p_k$ to denote the branching matrix.
Here, to avoid confusions, we use $\hat{t}$ for the coordinates that
appear in the Gromov-Witten theory.} $\hat{t}^1_1$ dual to the first 
descendant of the K\"{a}hler class.  We do not need an extra 
genus-keeping parameter, because for simple 
covers of  a fixed target space with a given number of marked points
$r$, choosing the degree of the map fixes the genus of the source
Riemann surface uniquely.  For the purpose of finding a nice
generating function, it is thus convenient to treat $r$ and $n$ as
independent indices, with the requirement that they be both 
non-negative.

For $r=0$, our previous computations of the simple Hurwitz numbers need
to be modified as
	\beq
		{T_{n,h,0}\over n!} = \sum^{n}_{k=1} \frac{(-1)^{k+1}}{k} 
		\sum_{ \mbox{\tiny
        $\begin{array}{c} n_1+\cdots +n_k=n
        \\ n_i > 0\end{array}$}} \prod_{i=1}^{k} (n_i)!^{2h-2}
		\left[ \sum_{\gamma\in D_{n_i}} (f^{\gamma})^{2-2h} \right]\ .
	\eeq
Also, note that $N_{1,h,r} = \delta_{r,0}$.
Then, we have
	\begin{claim} The generalized ``partition function" $Z(h)=\exp
	(\Phi(h))$  for all $h$ is given by
		\beq
		Z(h) = 1 + q + \sum_{q\geq 2} \sum_{\gamma\in {\SR_n}} \left(
		\frac{n!}{f^{\gamma}}\right)^{2h-2} \cosh \left[ 
		\left(\hspace{-1mm}\begin{array}{l}
		n\\2  \end{array}\hspace{-1mm}\right)
		\frac{\chi_{\gamma}(2)}{f^{\gamma}} \,\lambda
		\right] q^n . \label{eq:Z (h)}
		\eeq
	\end{claim}
\noindent
{\sc Proof:}  The proof is exactly the same as that of 
Claim~\ref{claim:dannazione}.  One just needs to keep tract of
extra factors in the general form of $B_{n,h,r}$.  For genus $h=0$,
when applying the Riemann-Hurwitz formula,
we must remember to use the correctly
 defined  arithmetic genus of reducible curves
and, as a result,
sum over all even $r\geq 0$ in $B_{n,0,r}$; doing so takes into
account
 the degree-1 covers in the exponential.  {\hfill
$\qed$}

%%%%%%%%%%%%%#################################################
%    
%%%%%%%%%%%%%#################################################
\setall
\section{Hodge Integrals on $\overline{\cM}_{g,2}$ and Hurwitz
Numbers} \label{sec:hodge}

In the modern language of Gromov-Witten theory, the simple Hurwitz 
numbers are equal to
	\beq
		\mu^{g,n}_{h,n} := \frac{T_{n,h,r}}{n!}=
		\langle \tau_{1,1}^{r}\rangle_{g,n} \ ,
	\eeq
where $r = 2(1-h)n + 2(g-1)$ and $\tau_{k,1}$ is the
$k$-th descendant of the K\"{a}hler class of the target 
genus-$h$ Riemann surface. 
We can organize these numbers into a 
generating function as follows:
	\beq
	H^g_h := \sum_{n} \frac{1}{r!} \langle \tau_{1,1}^{r}
	    \rangle_{g,n}
		\,(\hat{t}^1_1)^r \, e^{n\hat{t}} = \sum_{n} \frac{1}{r!} 
		\frac{T_{n,h,r}}{n!}
		\,(\hat{t}^1_1)^r \, e^{n\hat{t}} \ , \label{eq:Hgh}
	\eeq
where $\hat{t}^1_1$ and $\hat{t}$ are coordinates dual to $\tau_{1,1}$ 
and $\tau_{0,1}$, respectively.  In this paper, we have determined
\eqr{eq:Hgh} for all $g$ and $h$ up to degree $n=7$.  

For $h=0$ and $h=1$, these generating functions arise as genus-$g$ 
free energies of string theory on $\P^1$ and an elliptic curve
as target spaces, respectively, evaluated by setting all 
coordinates to zero except for $\hat{t}^1_1$ and $\hat{t}$.
For definitions of Hodge integrals, see \cite{FP,FP2}.

%%%%%%%%%%%%%%%
%
%%%%%%%%%%%%%%%
\subsection{Generating Functions for Hodge Integrals}
The Hurwitz enumeration problem has been so far investigated
intensely mainly for branched covers of the Riemann sphere. 
In this case, the almost simple Hurwitz numbers for covers 
with one general branch point can be expressed explicitly in 
terms of certain Hodge integrals.  An interesting application
of this development is to use the generating function 
for Hurwitz numbers
$\mu^{g,d}_{0,1} (d)$ to  derive a generating function
for Hodge integrals over the moduli space 
$\overline{{\cal M}}_{g,1}$.  More precisely, the formula
	\beq	\label{eq:F(t,k)}
	F(t,k) :=	1+ \sum_{g\geq 1}\, t^{2g} \sum_{i=0}^{g} \, k^i
		\int_{\overline{{\cal M}}_{g,1}} \psi^{2g-2+i} \,
			\lambda_{g-i} = \left( \frac{t/2}{\sin (t/2)}
			\right)^{k+1}
	\eeq
which was first obtained by Faber and Pandharipande in \cite{FP}
by using virtual localization  techniques has been rederived by 
Ekedahl {\it et al.} in \cite{ELSV} by using the generating
function for Hurwitz numbers for branched covers whose
only non-simple branch point has order equal to the degree of the
cover.

In this paper, we speculate a possible connection between the
Hurwitz numbers for $\P^1$ and generating functions for Hodge 
integrals on $\overline{{\cal M}}_{g,n}, n\geq 1$.  For this 
purpose, let us rewrite $F(t,k)$ as
	\beq \label{eq:new F}
	F(t,k) =	1+ \sum_{g\geq 1}\, t^{2g}\, \sum_{i=0}^{g} \, k^i
		\int_{\overline{{\cal M}}_{g,1}} 
		\frac{\lambda_{g-i}}{1-\psi_1} \ . 
	\eeq
Now,
recall that the simple Hurwitz numbers $\mu^{g,n}_{0,n} (1^n)$,
henceforth abbreviated $H_{g,n}$, have the following Hodge
integral expression 
\cite{FanP}:
	\beq \label{eq:H_{g,n}}
	H_{g,n} := \mu^{g,n}_{0,n} = \frac{(2g-2 +2n)!}{n!} 
		\int_{\overline{{\cal M}}_{g,n}} \frac{1-\lambda_1 + \cdots + 
		(-1)^g \lambda_g}{\prod^n_{i=1} (1-\psi_i)}
	\eeq
for $(g,n) \neq (0,1), (0,2)$.  The degree-1 simple Hurwitz numbers
are $H_{g,1} = \delta_{g,0}$, thus  \eqr{eq:H_{g,n}} yields the relation
	\beq
	\int_{\overline{{\cal M}}_{g,1}} \frac{1-\lambda_1 + \cdots
		+ (-1)^g \lambda_g}{1-\psi_1} =0, \ \mbox{ for $g\geq 1$.}
	\eeq
which implies from \eqr{eq:new F} 
that $F(t,-1) =1$, in accord with the known answer \eqr{eq:F(t,k)}.  
Naively, we thus see that the simple Hurwitz numbers are coefficients
of $F(t,k)$ evaluated at special $k$.

In a similar spirit, we can speculate a crude generating 
function for Hodge integrals with two marked points:
	\beq
		G(t,k) = \frac{1}{2} + \sum_{g\geq 1}\, t^{2g}\,
\sum_{i=0}^{g} \, k^i 
		\int_{\overline{{\cal M}}_{g,2}} 
		\frac{\lambda_{g-i}}{(1-\psi_1)(1-\psi_2)} \ . 
	\eeq
Our goal is to find a closed-form expression for this generating
function $G(t,k)$.  Without much work, we can immediately
evaluate $G(t,k)$ at certain special values of $k$:

\begin{claim} 	The generating function $G(t,k)$ can be evaluated at
$k=-1$ to be
       \beq \label{eq:hint1}
		G(t,-1) = \frac{1}{2} - \frac{1}{t^2} \left(
		\cos t + \frac{t^2}{2} -1
		\right)  = {1\over 2} \left( \frac{\sin (t/2)}{t/2}
\right)^2\ ,
	\eeq
and similarly at $k=0$ to be
	\beq  \label{eq:hint2}
		G(t,0) = \frac{1}{2} \left( \frac{t}{\sin t}\right) =
                \frac{1}{2}\, \frac{ t/2}{\sin (t/2)}\, {1\over \cos
(t/2)} \, . 
	\eeq
\end{claim}

\noindent
{\sc Proof:}
At $k=-1$, we can use \eqr{eq:H_{g,n}} to get
\beq \label{eq:G(t,-1)}
		G(t,-1) = \sum_{g\geq 0} \, (-1)^g 
		\frac{2 \,t^{2g}}{(2g+2)!} H_{g,2} \ .
	\eeq
We have previously computed $H_{g,2} = N_{2,0,2g+2}/2 = 1/2$, and  
we can then perform the summation in \eqr{eq:G(t,-1)} and get the 
desired result.
To evaluate $G(t,0)$, we use the following
$\lambda_g$-conjecture, which has been
recently proven by Faber and Pandharipande \cite{FP2}:
	\beq
	\int_{\overline{{\cal M}}_{g,n}} \psi^{\alpha_1}_1 \cdots
	\psi^{\alpha_n}_n \, \lambda_g = \left(\hspace{-1mm}
	\begin{array}{c}
		2g+n-3\\ \alpha_1, \ldots, \alpha_n 
		\end{array}\hspace{-1mm}\right) \, 
		\int_{\overline{{\cal M}}_{g,1}} \psi^{2g-2}_1 \,
\lambda_g \, . 
	\eeq
One can now compute
\barray
\int_{\overline{{\cal M}}_{g,2}}
\frac{\lambda_g}{(1-\psi_1)(1-\psi_2)} &=& {(2^{2g-1} -1)
\over (2g)!}\, | B_{2g}|
\earray
and obtain the result. {\hfill $\qed$}

To extract the terms without $\lambda_k$ insertions,
consider the scaling limit
	\barray 
	G( t\, k^{1\over 2}, k^{-1} ) &=& \frac{1}{2} + \sum_{g\geq 1}\, t^{2g}\,
\sum_{i=0}^{g} \, k^{g-i} 
		\int_{\overline{{\cal M}}_{g,2}} 
		\frac{\lambda_{g-i}}{(1-\psi_1)(1-\psi_2)} \\
	&\stackrel{k\ra 0}{\longrightarrow} & \frac{1}{2} + \sum_{g\geq 1}\, t^{2g}\,
\sum_{i=0}^{g} \,
		\int_{\overline{{\cal M}}_{g,2}} 
		\frac{1}{(1-\psi_1)(1-\psi_2)}. \label{eq:Glimit}
	\earray
The asymptotic behavior \eqr{eq:Glimit} can be explicitly evaluated
as follows:
	\begin{claim}
The asymptotic limit of $G(t,k)$ is
	\beq
	G( t\, k^{1\over 2}, k^{-1} )\,\stackrel{k\ra
	0}{\longrightarrow}\, {\exp \left(t^2 /3\right) \over 2\, t} 
        \sqrt{{\pi }}\, \mbox{\em Erf} \,\left[ \frac{t}{2} \right] \ ,
	\eeq
and thus, the integrals can be evaluated to be
\barray
\int_{\overline{{\cal M}}_{g,2}} \frac{1}{(1-\psi_1)(1-\psi_2)} &=&
{1 \over 2} \sum_{m=0}^g {1\over m! \, 12^m}\, { (g-m)! \over
(2g-2m+1)!} \ .
\earray
	\end{claim}

\noindent
{\sc Proof}: This is an easy consequence of the following 
Dijkgraaf's formula which appeared in the work of Faber \cite{Faber}:
	\beq \label{eq:Dijk-Faber}
	\langle \tau_0  \tau (w) \tau (z)\rangle = 	
	\exp \left( \frac{(w^3 + z^3 )\hbar }{24}\right) 
	\sum_{n\geq 0} \frac{n!}{(2n+1)!} \left[ {1\over 2}
	wz (w+z) \hbar\right]^n 
	\eeq
where $\tau (w) = \sum_{n\geq 0} \tau_n w^n$ and $\hbar$ is a formal
genus-expansion parameter defined by
	\beq
	 \langle\hspace{5mm} \rangle = \sum_{g \geq 0} \langle
	\hspace{5mm} \rangle_g \, \hbar^g\ .
	\eeq
Setting $w=z = \hbar^{-1} =t$ in \eqr{eq:Dijk-Faber} and noting that
	\beq
	\sum_{n\geq 0} \, \frac{1}{(2n+1)!!}\, t^{2n+1} = e^{t^2/2}
\,\sqrt{\frac{ \pi}{2}}\, \mbox{Erf} \left[ \frac{t}{\sqrt{2}}\right]
	\eeq
gives the result, upon using the string equation on the left-hand side.
{\hfill $\qed$}

For future reference, it would be desirable to find an explicit series 
expansion of $G(t,k)$.
Using Faber's Maple program for computing the intersection numbers on 
$\overline{M}_{g,n}$ \cite{Faber2}, the generating function can be seen to have an
expansion of the form
\barray \label{eq:GFaber}
G(t,k) &=&  {1\over 2} + \left({1\over12} + {1\over8} k\right)t^2 +
\left({7\over720} +
           {73\over 2880}k + 
          {49\over2880}k^2 \right) t^4 + \nonumber \\ \nonumber
&&  + \left({31\over 30240} + {253\over 72576} k + {983\over 241920} k^2 + 
          {1181 \over 725760} k^3 \right)t^6 + \\
&&  + \left({127\over 1209600} +
           {36413\over 87091200}k + 
           {37103\over 58060800} k^2 + {38869 \over 87091200}k^3 +
           {467 \over 3870720} k^4 \right) t^8 +\nonumber \\
&&  +  \left( {73 \over 6842880} + {38809\over 821145600} k + {122461\over
           1437004800}  k^2 +
           {86069\over 1094860800} k^3 + \right. \nonumber \\
&&   \hspace{1cm} + \left. {53597\over 1437004800} k^4 +
           {33631\over 4598415360} k^5\right) t^{10} + {\cal O}(t^{12}).
\earray

%%%%%%%%%%%%%
%
%%%%%%%%%%%%%

\subsection{Relation to Hurwitz Numbers $\mu^{g,2k}_{h,2}(k,k)$}
We now relate the generating function $G(t,k)$ to the Hurwitz numbers
$\mu^{g,2k}_{h,2}(k,k)$, which we are able to compute explicitly.  This
connection allows us to evaluate $G(t,k)$ for all $k \in \QZ_{<0}$.
From the work of \cite{ELSV}, we know that
	\beq
	\mu^{g,2k}_{0,2} (k,k) = { (2k+ 2g)!\over 2} {k^{2k}\over
	(k!)^2} \int_{\overline{M}_{g,2}} {c (\Lambda^{\vee}_{g,2})
	\over (1-k \psi_1) (1-k\psi_2)} ,
	\eeq
which we can rewrite as
	\beq
 	\mu^{g,2k}_{0,2} (k,k) ={ (2k+ 2g)!\over 2} {k^{2k+2g-1}\over
	(k!)^2}   \sum_{i=0}^g \, k^i\,
	\int_{\overline{M}_{g,2}} { (-1)^{g-i}\,\lambda_{g-i}
	\over (1- \psi_1) (1-\psi_2)} \,.
	\eeq
This implies that for integers $k>0$,
	\beq \label{eq:Gmu}
	G(it, -k) = {1\over 2} + \sum_{g\geq 1}  {2\, t^{2g}\over (2k+2g)!}\,
	{(k!)^2 \over k^{2k+2g-1}}\, \mu^{g,2k}_{0,2}(k,k)\, .
	\eeq
By using the expansion \eqr{eq:GFaber} and matching coefficients with
\eqr{eq:Gmu}, one can thus obtain the
Hurwitz numbers	$\mu^{g,2k}_{0,2} (k,k)$.  We have listed the numbers 
for $g \leq 6$ in Appendix~\ref{sec:app mu2k}.

It is in fact possible to determine the Hurwitz numbers
$\mu^{g,2k}_{0,2}(k,k)$ from the work
of Shapiro {\it et al.}\/
on enumeration of edge-ordered graphs \cite{SSV}.  According to
theorem 9 of their paper\footnote{Actually, their formula has a minor
mistake for the case when $n=2k$ is partitioned into
$(k,k)$ for odd genus.  More precisely,
when the summation variable $s$ in their 
formula equals $(g+1)/2$, for an odd genus $g$, there is a symmetry 
factor of $1/2$ in labeling the edges because the two disconnected
graphs are identical except for the labels.}, 
	the Hurwitz numbers $\mu^{g,2k}_{0,2} (k,k)$ are 
		given by
			\barray
				\mu^{g,2k}_{0,2} (k,k) &=& N(2k,2k+2g,
		(k,k)) -  
				{ 2k \choose k} \frac{(2k
		+2g)!}{(2k)!}\,  k^{2k-2+2g} \times \nonumber\\
		&& \ \times\, {1\over 2}\,
\left[ \sum_{s=0}^{g+1} \, \delta^k_{2s}\,
		\delta^k_{2g+2-2s} \right],
%				&& \ \times 
%			\left[ \sum_{s=0}^{\lfloor {g-1\over 2} \rfloor}
%  \, \delta^k_{2s}
%                        \, \delta^k_{2g+2-2s} + \frac{3+(-1)^{g}}{4}\,
%                        \delta^k_{2\lfloor\frac{g+1}{2}\rfloor}\, 
%                        \delta^k_{2g+2-2\lfloor\frac{g+1}{2}\rfloor}\right]
			\label{eq:claim N}
			\earray
where the numbers $\delta^k_{2g}$ are defined by
			 \[ 
			 \sum^{\infty}_{g=0} \, \delta^k_{2g} \,
		t^{2g} = \left( 
			 {\sinh (t/2) \over t/2 }\right)^{k-1} \, 
			 \]
and can be written explicitly as
\[
\delta^k_{2g} = {1 \over (k+2g-1)!} \sum_{m=0}^{k-1} {k-1 \choose m}
(-1)^m \left( {{k-1\over 2}-m}\right)^{k+2g-1} \ .
\]
The number $N(2k,2k+2g,(k,k))$, which counts the number of certain
edge-ordered graphs, is given by
\begin{equation}
N(2k,2k+2g,(k,k)) \ = \ {\left| C(k,k) \right| \over [(2k)!]^2}
\sum_{\rho \vdash 2k} f^{\rho} (h(\rho')-h(\rho))^{2k+2g} \chi_{\rho}(k,k),
\label{eq:def N}
\end{equation}
where $\left| C(k,k) \right|$ is the order of the conjugacy class
$C(k,k)$, $\rho'$ is the partition conjugate to $\rho$, and
$h(\rho)=\sum^m_i (i-1)\rho_i$ for $\rho=(\rho_{\mbox{\tiny
$1$}},\ldots,\rho_m) \vdash 2k$.  Hence, the problem of finding
$\mu^{g,2k}_{0,2} (k,k)$ reduces down to evaluating \eqr{eq:def N}.

	\begin{claim}  For $k \geq 2$, 
			\begin{eqnarray}
			N(2k,2k+2g,(k,k)) &=&
\frac{(k-1)!}{2\,k\cdot k!\,\left(2\,k \right)!} 
\,\left\{ \frac{2\, 
         {\left[k\, (k -2) \right] }^{2\,g + 2\,k}\,\left( 2\,k
         \right) !}{k!\, 
         \left( 1 + k \right) !}\, + \right. \nonumber \\
&&       + \ \sum_{m = 0}^{k - 1}{{2\,k - 1}\choose m}\,{\left( -1 \right)
}^m\,
         {\left[ k\, (2k-2m-1) \right]
         }^{2\,k + 2\,g} + \nonumber \\
&&       + \  \sum_{m = k}^{2k - 1}{{2\,k - 1}\choose m}\,{\left( -1
			\right) }^{m-1}\,
         {\left[ k\, (2k-2m-1) \right]
         }^{2\,k + 2\,g} + \nonumber \\
&&  \left.     +\ 2 \sum_{m = 0}^{k - 3}\sum_{p = 1}^{k - m - 1}
          {k-1 \choose m} {k-1 \choose m+p}\right. \times \nonumber\\
	&&\ \ \ \times\left.
          {p^2 \over k^2-p^2}\, {(2k)! \over (k!)^2}
          \, {\left( -1 \right) }^{p + 1}
           {\left[ k\,\left( k - 2\,m - p - 1 \right)  \right] }^{2\,k
         + 2\,g} \right\}. \nonumber
			\end{eqnarray}
			 \end{claim}
\noindent
{\sc Proof:}  To each irreducible representation
labeled by $\rho=(\rho_{\mbox{\tiny $1$}},\ldots,\rho_{\mbox{\tiny $j$}}) \vdash 2k$,
we can associate a Young diagram with $j$ rows, the $i^{th}$ row
having length $\rho_{i}$.  According to the {\em
Murnaghan-Nakayama} rule, the diagram corresponding to an
irreducible representation $\rho$ for which $\chi_{\rho}(k,k)\neq 0$,
must be either $(a)$ a hook or $(b)$ a union of two hooks.  After
long and tedious computations, we arrive at the following results: \\
%%%%%%%%%%%%%%%%%%%%%%%
$(a)$ There are $2k$ ``one-hook'' diagrams. \\
%%%
\mbox{\hspace{.5cm}} $(i)$ The diagram with leg-length $m$ for $0 \leq m \leq
k-1$ gives
\[ f^\rho = {2k-1 \choose m} \ , \ \chi_\rho(k,k) = (-1)^m \ , \ 
h(\rho')-h(\rho) = k (2k-2m-1).\] \\
%%%
\mbox{\hspace{.5cm}}$(ii)$ The diagram with leg-length $m$ for $k \leq m \leq 2k-1$ gives
\[ f^\rho = {2k-1 \choose m} \ , \ \chi_\rho(k,k) = (-1)^{m-1} \ , \ 
h(\rho')-h(\rho) = k (2k-2m-1).\] \\
%%%%%%%%%%%%%%%%%%%%%%%
$(b)$ There are $k(k-1)/2$ ``two-hook'' diagrams.\\
%%%
\mbox{\hspace{.5cm}}$(i)$ For each value of $m$ and $p$ satisfying
$0\leq m\leq k-3$ and $1\leq p \leq k-m-1$, respectively, there is a
diagram with $k-m$ columns and 
$p+m+1$ rows.  Such diagram has\\
\[ f^\rho = {k-1 \choose m} {k-1 \choose m+p} {p^2 \over k^2 -p^2}
{(2k)! \over (k!)^2}\, , \]
\[ \chi_\rho(k,k) = 2 (-1)^{p+1} \ , \ h(\rho')-h(\rho) =
k(k-2m-p-1)\, .
\]
%%%
\mbox{\hspace{.5cm}}$(ii)$ One diagram has $2$ columns and $k$ rows.
It corresponds to the irreducible representation with
\[f^\rho = {(2k)!\over k!(k+1)!} \ , \ \chi_\rho(k,k) = 2 \ , \ 
h(\rho')-h(\rho) = k(k-2)\, .
\]
%%%%%%%%%%%%%%%%%%%%%%
Furthermore, after some simple combinatorial consideration, we find
that  $|C(k,k)| = $ \break $(2k)!(k-1)!/(2k\cdot k!)$.
Finally, substituting in \eqr{eq:def N} the  values
of $f^\rho, \chi_\rho(k,k)$ and $h(\rho')-h(\rho)$ for the above
$k(k+3)/2$ irreducible representations gives the desired result.
\  \hfill $\qed$

By using \eqr{eq:Gmu} and \eqr{eq:claim N}, we can now rewrite
$G(it,-k)$ as

\begin{claim} For integral $k \geq 2$,
	\barray  
	G(it, -k) &=&  {2\,(k-1)!\over (k+1)! \, t^{2k}} \cosh[(k-2)t] 
	 + \,  {2\,(k!)^2\over  k\,(2k)!\, t^{2k}} 
	\sum_{m=0}^{k-1} {2k-1\choose m} (-1)^m  \cosh[(2k-2m-1) t] \nonumber\\
	&& +\,
	{2\over k \, t^{2k}} \sum_{m=0}^{k-3} \sum_{p=1}^{k-m-1} {k-1\choose m} 
	{k-1 \choose m+p} 
	{p^2 \over k^2 - p^2} (-1)^{p+1} \cosh[(k-2m-p-1)t]  \nonumber\\
	&&-\,
	{1\over k\, t^2} \left( 
	{\sinh[ t/2] \over (t/2)}\right)^{2k-2}\, .
	\earray \label{eq:Gitk}
\end{claim}

\noindent
{\sc Proof:} By substituting the expression \eqr{eq:claim N} into 
\eqr{eq:Gmu} and summing over the $\delta$ terms, we get
	\barray
	G(it, -k) &=&  {1\over 2}  + \sum_{g\geq 1} {2\, (k!)^2 \, 
	t^{2g} \over
	(2k+2g)!\, k^{2k+2g-1} } \, N(2k,2k+2g, (k,k))\nonumber\\
	&&\  -\, {1\over kt^2} 
	\left( \sinh (t/2) \over t/2 \right)^{2k-2} + 
	{\delta^k_0\delta^k_0\over kt^2} + 
	{2\over k}\, \delta^k_0 \delta^k_2\nonumber\\ 
    &=&  {1\over 2}  + \sum_{\ell\geq 0} {2\, (k!)^2 \, 
	t^{2\ell-2k} \over
	(2\ell)!\, k^{2\ell-1} } \, N(2k,2\ell, (k,k))
	- \sum_{\ell\geq 0}^k {2\, (k!)^2 \, 
	t^{2\ell-2k} \over
	(2\ell)!\, k^{2\ell-1} } \, N(2k,2\ell, (k,k))\nonumber\\
	&&\  -\, {1\over kt^2} 
	\left( \sinh (t/2) \over t/2 \right)^{2k-2} + 
	{\delta^k_0\delta^k_0\over kt^2} + 
	{2\over k}\, \delta^k_0 \delta^k_2\, . \nonumber\\
	\earray
But, by Lemma~\ref{lemma:ssv1}, $N(2k,2\ell, (k,k))=0$ for $\ell
\leq k-2$.  Furthermore, we have
	\[
	-\,{2 (k!)^2 \over (2k)!\, k^{2k-1}} N(2k,2k,(k,k)) + 
	{2\over k}\, \delta^k_0 \delta^k_2 =-\,
	{2 (k!)^2 \over (2k)!\, k^{2k-1}} \mu^{0,2k}_{0,2}(k,k) 
	= -\,{1\over 2}\,, 
	\]
and
	\[
	-\,{2 (k!)^2 \over (2k-2)! k^{2k-3} t^2} N(2k,2k-2,(k,k)) + 
	{\delta^k_0\delta^k_0\over kt^2} \propto N_c(2k,2k-2, (k,k)) =0,
	\]
where we have used the known fact \cite{SSV} that 
    \[ 
	\mu^{0,2k}_{0,2}(k,k) = {2k \choose k} {k^{2k-1}\over 4}
	\]
and Lemma~\ref{lemma:ssv2}.
Thus, we have
	\beq
	G(it,-k)=   \sum_{\ell\geq 0} {2\, (k!)^2 \, 
		t^{2\ell-2k} \over
		(2\ell)!\, k^{2\ell-1} } \, N(2k,2\ell, (k,k))
		 -\, {1\over kt^2} 
		\left( \sinh (t/2) \over t/2 \right)^{2k-2} ,
	\eeq
where the first term can now be easily summed to yield our claim.
{\hfill $\qed$}

It turns out  that there are some magical 
simplifications,
and we find for a few low values of $k$ that
	\barray
	G(t, -1) &=& \frac{1}{2}\, 
		 \left(\sin (t/2) \over t/2 \right)^2, \nonumber\\
	G(t, -2) &=& \frac{1}{6}\, \left[ 2+  \cos (t)  \right] 
	   \left(
	 \sin (t/2) \over t/2 \right)^4 ,\nonumber\\
	G(t,-3) &=& \frac{1}{30}\, \left[ 8+ 6\, \cos (t) + \cos (2t) \right] 
	\left(
	 \sin (t/2) \over t/2 \right)^6 ,\nonumber\\
    G(t,-4) &=& \frac{1}{140}\, \left[ 32+ 29\, \cos (t) + 
	 8 \,\cos (2t) + \cos (3t) \right] 
	 	\left(
	 \sin (t/2) \over t/2 \right)^8 ,\nonumber\\
    G(t,-5) &=& \frac{1}{630}\, \left[ 128+ 130\, \cos (t) + 
	 	 46 \,\cos (2t) + 10\, \cos (3t)+ \cos (4t) \right] 
	 	 	\left(
	 	 \sin (t/2) \over t/2 \right)^{10} ,\nonumber\\
     G(t,-6) &=& \frac{1}{2772}\, \left[ 512 + 562\, \cos (t) + 
	 	 232 \,\cos (2t) + 67\, \cos (3t)+ 12\, \cos (4t)\right. \nonumber\\
		 &&\left. \hspace{2cm} + \,\cos(5t) \right]
	 	 	\left(
	 	 \sin (t/2) \over t/2 \right)^{12} ,\nonumber\\
     G(t,-7) &=& \frac{1}{4(3003)}\, \left[ 2048 + 2380\, \cos (t) + 
	 	 1093 \,\cos (2t) + 378\, \cos (3t)+ 92\, \cos (4t)\right. \nonumber\\
		 &&\left. \hspace{2cm} + \, 14 \, \cos(5t)+ \cos(6t) \right] 
	 	 	\left(
	 	 \sin (t/2) \over t/2 \right)^{14} ,\nonumber\\ 
	\earray
and so forth.
We have explicitly computed $G(t,-k)$ for $k\leq 60$, and based on these 
computations, we conjecture the following general form:	
	\begin{conjecture}\label{conj:G}
 For integers $k\geq 1$, the generating
function is given by 
\barray
  G(t,-k)&=&
    \frac{2(k-1)!\,k!}{(2k)!}\left(\frac{\sin(t/2)}{t/2}\right)^{2k} 
    \left[ 2^{2(k-2)+1}
    +\sum_{n=1}^{k-1} \left[\sum_{i=0}^{k-n-1} {2k-1 \choose i}\right]
	\cos(nt)\right].\nonumber\\ \label{eq:conj G}
\earray
	\end{conjecture}
Let us rewrite the summation as follows:
	\barray
	\sum_{n=1}^{k-1} \left[\sum_{i=0}^{k-n-1} {2k-1 \choose i}\right]
	\cos(nt) &=& \sum_{\ell =0}^{k-2} {2k-1 \choose \ell}
	\left( \sum_{n=1}^{k-1-\ell} \,\cos (nt)\right)\nonumber\\
	&=& {1\over 2}\sum_{\ell =0}^{k-2} {2k-1 \choose \ell}
	\left[ {\sin \left[( 2k-1-2\ell) t/ 2\right]\over \sin
	(t/2)} -1 \right]  . \label{eq:simple}
	\earray
The last expression in \eqr{eq:simple} can now be explicitly summed,
	leading to an expression which can be analytically continued
	to all values of $k$. 
	After some algebraic manipulations, we obtain the following
	corollary to Conjecture~\ref{conj:G}:
	\begin{conjecture}  For all\footnote{For $k$ non-positive
integers and half-integers, the below expression of $G(t,-k)$ appears
to be divergent.  For these cases, one might try first expanding
$G(t,-k)$ 
in $t$ and setting $k$ equal to the desired values.} $k$, the generating
function as a 
formal power series in $\Q [k] [[t]]$ is given by
\barray
     G(t,-k)&=& {2^{2k -1} \over \sqrt{\pi }}\, \frac{
\Gamma(k)\,\Gamma(\frac{1}{2} + k)}{\Gamma(2k+1)}
\left(\frac{\sin(t/2)}{t/2}\right)^{2k}  \,
         {1 \over \sin(t/2)} \,
           \nonumber\\
      &&   \ \times\, \left[   \sin(t/2) +  \Re \left(
           i\, e^{i \,t/2}\,
             \mbox{}_2F_1(1,-k,k,-e^{-i \,t})\right)\right] \,
			 , \label{eq:G hyper}
\earray
where $\Re$ denotes the real part.
	\end{conjecture}
We have checked that our conjectural formula \eqr{eq:G hyper} indeed
reproduces all the terms in \eqr{eq:GFaber}.

%%%%%%%%%%%%%%
%
%%%%%%%%%%%%%%

\subsection{Possible Extensions}

Motivated by our results, let us consider  a similar generating
function for the case of more marked points:
	\beq
	 G_n (t,k) := {n!\over (2n-2)!}H_{0,n} + \sum_{g\geq 1}\, t^{2g}\,
\sum_{i=0}^{g} \, k^i  \int_{\overline{{\cal M}}_{g,n}} 
		\frac{\lambda_{g-i}}{(1-\psi_1)\cdots(1-\psi_n)}.
	\eeq
At $k=-1$, it can be evaluated in terms of simple Hurwitz numbers as
	\beq
	G_n(t,-1) = n! \sum_{g=0}^{\infty} {(-1)^g\,H_{g,n}\over (2g
+2n-2)!}\,t^{2g}.
	\eeq
Interestingly, our previous generating function for simple Hurwitz
numbers \eqr{eq:Phi}, with $\lambda = it$, 
is related to $G_n(t,-1)$:
	\beq
	\Phi (0) \mbox{\Large $|$}_{\lambda=it} 
=  \log Z(0)=\sum_{n\geq 1} {(it)^{2n-2}\over n!} \, G_n
(t,-1)\, q^n\,.
	\eeq
Hence, we have
	\beq
	G_n(t,-1) = {n!\over t^{2n-2}} \,\sum^n_{k=1}
{(-1)^{k-n} \over n} \sum_{ \mbox{\tiny
        $\begin{array}{c} m_1+\cdots +m_k=n
        \\ m_i > 0\end{array}$}} W_{m_i}\cdots W_{m_k} 
	\eeq
where $W_1 =1$ and
	\beq
	W_{m_i} = \sum_{\gamma\in {\SR_{m_i}}}\left(
		\frac{f^{\gamma}}{m_i!}\right)^{2} \cos \left[ 
		\left(\hspace{-1mm}\begin{array}{c}
		m_i\\2  \end{array}\hspace{-1mm}\right)
		\frac{\chi_{\gamma}(2)}{f^{\gamma}} \,t
		\right]\, .
	\eeq
This relation might suggest a possible connection between
the symmetric group $S_n$ and the geometry of
the moduli space of marked Riemann surfaces.

Of course, $G_n(t,-1)$ can be also explicitly computed from our previous
computations 
of the simple Hurwitz numbers $H_{g,n}$.
  For example, we find that
	\barray
	G_3(t,-1) &=& {(2+\cos(t)) \over 3}
		\left( {\sin(t/2)\over t/2} \right)^4,\nonumber\\
	G_4(t,-1) &=&  \frac{\left( 20 + 21\,\cos (t) + 6\,\cos
      (2\,t)  + \cos (3\,t) \right)}{12} \,
       \left({\sin (t/2)\over t/2}\right)^6\, , \nonumber\\
       G_5(t,-1) &=& \left({\sin (t/2)\over t/2}\right)^8  {1\over 60} 
        \left[ 422 + 608\,\cos (t) + 305\,\cos (2\,t) +
        \right. \nonumber\\ 
      &&\hspace{1cm}\left. +
      120\,\cos (3\,t) + 36\,\cos (4\,t) +  
      8\,\cos (5\,t) + \cos (6\,t) \right], \nonumber\\
	G_6(t,-1) &=& \left({\sin(t/2)\over t/2}\right)^{10} {1\over
360}\times  
\left[ 16043 + 26830\,\cos (t) + 17540\,\cos (2\,t)+\right. \nonumber \\
 &&\hspace{1cm}  + 9710\,\cos
(3\,t) +     4670\,\cos (4\,t) + 1966\,\cos (5\,t) + 715\,\cos
(6\,t) +\nonumber\\ 
&&\hspace{1cm}+ 220 \,\cos (7\,t) +  \left. 
      55\,\cos (8\,t) + 10\,\cos (9\,t) + \cos (10\,t) \right]\, .
	\earray
Similarly, $G_n(t,0)$ can be computed by using the
$\lambda_g$-conjecture.  For example, one can easily show that
	\beq
	G_3(t,0) = {(3t/2)\over \sin (3t/2)}\, ,
	\eeq
{\it et cetera}.
Although we are able to compute the generating function $G_n(t,k)$ at
these particular values, it seems quite difficult--nevertheless 
possible--to determine its closed-form expression for all $k$.  It
would be a very intriguing project to search for the answer.

%%%%%%%%%%%%%################################################
%    
%%%%%%%%%%%%%################################################
\setall
\section{Conclusion, or An Epilogue of Questions Unanswered}
To recapitulate, the first part of our paper studies the simple 
branched covers of Riemann surfaces by Riemann surfaces of 
arbitrary genera.  Upon fixing the degree of the irreducible
covers,  we have obtained closed form answers for simple
Hurwitz numbers for arbitrary
source and target Riemann surfaces, up to degree 7. 
For higher degrees, we have given a general prescription for extending
our results.
Our computations are novel in the sense that the previously known
formulas fix the genus of the source and target curves 
and vary the degree as a free parameter.  Furthermore, by
relating the simple Hurwitz numbers to descendant Gromov-Witten
invariants, we have obtained the explicit generating functions 
\eqr{eq:Z (h)}
for the number of inequivalent reducible covers for arbitrary
source and target Riemann surfaces.  For an elliptic curve
target, the generating function \eqr{eq:Z (1)} is known to be
a sum of
quasi-modular forms.  More precisely,  in the expansion 
	\beq
	Z = \sum_{n=0}^{\infty} 
	\, A_n (q) \, \lambda^{2n}\ ,
	\eeq
the series $A_n(q)$ are known to be
quasi-modular of weight $6n$ under the full 
modular group $PSL(2,\QZ)$. 
 Our general answer \eqr{eq:Z (h)}
for an arbitrary target genus differs from the elliptic curve
case only by the prefactor $(n!/f^{\gamma})^{2h-2}$.  Naively,
it is thus tempting to hope that the modular property persists,
so that in the expansion
	\beq
	Z (h) = \sum_{n=0}^{\infty} \, A^h_n (q) \, \lambda^{2n},
	\eeq
the series $A^h_n (q)$ are quasi-automorphic forms, perhaps 
under a genus-$h$ subgroup of \break $PSL(2,\QZ)$.

Throughout the paper, we
have taken caution to distinguish two different conventions
of accounting for the automorphism groups of 
the branched covers and have clarified their relations
when possible.
The recent developments in the study of Hurwitz numbers involve
connections to the relative Gromov-Witten theory and Hodge integrals
on the moduli space of stable curves.   In particular, Li {\it et al.} 
have
obtained a set of recursion relations for the numbers $\mu^{g,n}_{h,w}
(\alpha)$  by applying the
gluing formula to the relevant relative GW invariants
\cite{Li}.  Incidentally, 
these recursion relations require as initial data the
knowledge of simple Hurwitz numbers, and
our work would be useful for applying the relations
as well.

Although we cannot make any precise statements at this stage,
our work may also be relevant to understanding the conjectured
Toda hierarchy and the Virasoro constrains for Gromov-Witten
invariants on $\P^1$ and elliptic curve. 
It has been shown in \cite{JSS} 
that Virasoro constraints lead to certain recursion relations
among simple Hurwitz numbers for a $\P^1$ target.  It might be
interesting to see whether there exist further connections
parallel to these examples.
The case of an elliptic curve target seems, however,
 more elusive at the moment.
The computations of the Gromov-Witten invariants for an elliptic curve
are much 
akin to those occurring for Calabi-Yau three-folds.  For instance, a
given $n$-point  
function receives contributions from the stable maps of all degrees,
in contrast to  
the Fano cases in which only a finite number of degrees yields the
correct dimension  
of the moduli space.  Consequently, the recursion relations and the
Virasoro constraints 
seem to lose their efficacy when one considers the Gromov-Witten 
invariants of an
elliptic curve. 
It is similar to the ineffectiveness of the WDVV equations for
determining the  
number of rational curves on a Calabi-Yau three-fold.

%%%%%%%%%%%%%%%%%%%%%%%%%%%%%%%
%
%%%%%%%%%%%%%%%%%%%%%%%%%%%%%%%
\vspace{1cm}
\noindent
{\bf Acknowledgments}

\noindent
We gratefully acknowledge Ravi Vakil for numerous valuable discussions and
suggestions.
J.S.S. thanks Prof. Gang Tian and Y.-H. He for discussions.

%%%%%%%%%%%%%######################################
%    APPENDIX
%%%%%%%%%%%%%######################################
\newpage
\appendix
\setall
\section{Rudiments of the Symmetric Group $S_n$}
\def\a#1{{\alpha_{#1}}}

It is well-known that the conjugacy classes, and thus the irreducible
representations,  of the symmetric group
$S_n$ are in one-to-one correspondence with
 distinct ordered partitions of $n$.
Let us consider an irreducible representation of $S_n$ labeled by 
the ordered 
partition $\gamma = (n_1, \ldots, n_m) \vdash n$ , where
$n_1 \geq n_2 \geq \ldots \geq n_m$.  
Let $p_i = n_i + m - i$ and define the Van der Monde determinant
	\begin{equation}
	D(p_1,\ldots,p_m) \equiv \left| 
 	\matrix{ p_1^{m-1} & p_1^{m-2} & \cdots & p_1 & 1 \cr
          p_2^{m-1} & p_2^{m-2} & \cdots & p_2 & 1 \cr
             \vdots &    \vdots & \vdots & \vdots & \vdots \cr
          p_m^{m-1} & p_m^{m-2} & \cdots & p_m & 1 \cr
        } \right|.
\end{equation}
Then, the irreducible characters evaluated at the conjugacy classes
$(1^n)$ and $(2)$ can be written
respectively as  
	\begin{equation} 
	\chi_{\gamma} \left( 1^n \right) =
	\frac{n!}{p_1!\ p_2!\ \cdots p_m!} 
	D(p_1,\ldots,p_m) 
	\end{equation}
and
	\begin{equation}
	\chi_{\gamma} \left( 2\right) =
	(n-2)!\ 
	\sum_{i\in {\cal I}} 
	\frac{D(p_1,\ldots,p_{i-1},p_i - 2,
	p_{i+1},\ldots,p_m)}{p_1!\cdots 
	p_{i-1}!\ (p_{i}-2)!\ p_{i+1}!\ \cdots p_m!}, 
	\end{equation}
where the index set ${\cal I}$ is defined as $\left\{i\in
\{1,\ldots,m\} \| (p_i -2)\geq 0\right\}$.
Furthermore, these irreducible characters satisfy the simple relation
  	\begin{equation} \label{eq:argument}
	{n \choose 2}\ \frac{\chi_{\gamma} \left(2
	\right)}{\chi_{\gamma} \left( 1^n \right)} \ = \ 
   	\frac{1}{2} \sum_{k=1}^{m} n_k (n_k + 1)
           -    \sum_{k=1}^{m} k \cdot n_k\, ,
	\end{equation}
which we utilize in the paper.

%%%%%%%%%%%%%%################################
%
%%%%%%%%%%%%%%################################

\setall
\section{Useful Facts} \label{app:details}
\begin{lemma} \label{lemma:1} Let $L = \langle e_1, e_2
\rangle := \QZ e_1 + \QZ e_2$ be a two-dimensional 
lattice generated by $e_1$ and $e_2$.  Then,
the number of inequivalent sublattices $L'\subset L$ of index  $[L : L'] =n$
is given by $\sigma_1(n) :=\sum_{d|n} d$.
\end{lemma}

\noindent
{\sc Proof:} Let $f_1 = d e_1\in L'$ be the smallest multiple of
$e_1$.  Then, there exists  
$f_2= a e_1 + b e_2 \in L', a< d,$ such that $L'$ is generated by
$f_1$ and $f_2$ over $\QZ$. 
It is clear that the index of this lattice is $db$.  Thus, for each
$d$ dividing the  
index $n$, we have the following $d$ inequivalent sublattices:
$\langle de_1, (n/d) e_2 
\rangle , 
\langle de_1,  e_1+ (n/d) e_2\rangle, \ldots , \langle de_1, (d-1) e_1
+ (n/d) e_2\rangle$.   
The lemma now follows.  {\hfill $\qed$}

%%%%%%%%

	\begin{lemma} \label{lemma:sigma_1}
		Let $\pi(m)$ be the number of distinct ordered partitions of a 
		positive integer $m$ into positive integers.  Then, the
		function $\sigma_1 (n)$ has the following expression:
			\beq
			\sigma_1 (n) = n \sum^n_{k=1} \frac{(-1)^{k+1}}{k}
			\sum_{m_1+\cdots m_k = n} \left(\prod^{k}_{i=1}\pi (m_i) 
			\right)\ .
			\eeq
	\end{lemma}
\noindent
{\sc Proof:}  As is well-known, the functions $\pi (m)$ arise as
coefficients of the expansion of $q^{1/24} \, \eta(q)^{-1}$, i.e.
	\beq  \label{q eta}
		\frac{q^{1/24}}{\eta (q)} = 1 + \sum^{\infty}_{m=1} 
			\pi (m) \, q^m \ .
	\eeq
We  take $\log$ of both sides of \eqr{q eta} and $q$-expand the resulting 
expression on the right hand side.  Now, using the fact that 
	\beq
		\log\left( q^{1/24}\eta (q)^{-1}\right) = \sum^{\infty}_{n=1}
				\frac{\sigma_1 (n)}{n} q^n \ , 
	\eeq
we match the coefficients of $q^n$ to get the desired result.
{\hfill $\qed$}

%%%%%%%%%%%%%%%
As in \cite{SSV}, let $N(n,m,\nu)$ be the number of edge-ordered graphs with 
$n$ vertices, $m$ edges, and $\nu$ cycle partition, and $N_c(n,m,\nu)$ the number
of connected such graphs.  Then,
	\begin{lemma} \label{lemma:ssv1}
		$N(2k, 2\ell, (k,k))=0$ for $\ell \leq k-2$.
	\end{lemma}
\noindent
{\sc Proof:} These constraints follow from Theorem 4 of \cite{SSV} which 
states that the length $l$ of the cycle partition must satisfy the 
conditions $c\leq l \leq \min(n,m-n +2c)$ and $l = m-n (\mod 2)$, where
$c$ is the number of connected components.  In our case, $l=2$ and the
second condition is always satisfied.  The first condition, however,
is violated for all $\ell \leq k-2$ because $c\leq 2$ and thus $\min
(2\ell -2k +2c) \leq 0$.
{\hfill $\qed$}

\vspace{2mm}
\noindent
Similarly, one has
	\begin{lemma} \label{lemma:ssv2}
		$N_c(2k, 2k-2, (k,k)) =0$.
	\end{lemma}
{\sc Proof:}
This fact again follows from Theorem 4 of \cite{SSV}.  Here, $c=1$ and
$\min (n,m-n+2c) =0$, whereas $\ell =2$, thus violating the first
condition of the theorem.
{\hfill $\qed$}

%%%%%%%%%%%%%%%%
%
%%%%%%%%%%%%%%%%

\setall
\section{Computation of Simple Hurwitz Numbers} \label{app:comp}
For computations of $\tilde{\mu}^{g,n}_{h,n} = N_{n,h,r}$, we
will need the following relation among the numbers of
irreducible and reducible covers \cite{Mednykh1}:
	\beq
	T_{n,h,\sigma} = \sum_{k=1}^n {(-1)^{k+1}\over k} 
		\sum_{ \mbox{\tiny
        $\begin{array}{c} n_1+\cdots +n_k=n
        \\ \sigma_1 + \cdots \sigma_k = \sigma \end{array}$}}
		{n \choose n_1, \ldots, n_k} \, B_{n_1,h, \sigma_1} 
		\cdots B_{n_k,h, \sigma_k} \label{eq:TB}
	\eeq
where
	\beq
	B_{n,h,r} = (n!)^{2h-1} \left(\hspace{-1mm}\begin{array}{l}
			n\\2 \end{array}\hspace{-1mm}\right)^r \, \left[	
			\sum_{\gamma\in {\SR_n}} \frac{1}{(f^{\gamma})^{2h-2}} \left(
			\frac{\,\chi_{\gamma}(2)}{f^{\gamma}}\right)^r
		\right].
	\eeq
Furthermore, in these computations, 
we assume that $r$ is positive unless indicated otherwise.
%%%%%%%%%%%%%%%%%
%  Degree 1 and 2
%%%%%%%%%%%%%%%%%

\subsection{Degree One and Two}
It is clear that the degree-one simple Hurwitz numbers are
given by
	\beq
	\mu^{g,1}_{h,1} (1) = \delta_{g,h} \ .
	\eeq

The number of simple double covers of a genus-$h$ Riemann surface by
by genus-$g$
Riemann surfaces can be obtained by using the work of Mednykh on
Hurwitz numbers for the case where all branchings have the order equal
to the degree of the covering \cite{Mednykh}.   

\begin{claim} \label{claim:2} The simple Hurwitz numbers
$\tilde{\mu}^{g,2}_{h,1}(2)$ are equal to $2^{2h}$ for
$g >2(h-1) + 1$.
\end{claim}

\noindent
{\sc Proof:} For $g > 2(h-1) +1$, the number $r$ of simple branch 
points is positive, and we can use the results of Mednykh
\cite{Mednykh}. 
 Let $p$ be a prime
number and $D_p$ the set of all irreducible representations of the
symmetric group $S_p$. Then, Mednykh shows that 
 the number $N_{p,h,r}$ 
of  inequivalent degree-$p$ covers of a
genus-$h$ Riemann surface by genus-$g$ Riemann with $r$ branch
points\footnote{The Riemann-Hurwitz formula determines $r$ to be
$r = [2(1-h) p + 2(g-1)]$.}
of order-$p$ is given by
	\beq
	N_{p,h,r} = \frac{1}{p!}\, T_{p,h,r} + p^{2h-2} [ (p-1)^r +
(p-1)(-1)^r] \, ,
	\eeq
where
	\beq
	T_{p,h,r} = p! \, \sum_{\gamma \in D_p} \left(
\frac{\chi_{\gamma}(p)}{p}\right)^r \, \left( \frac{p!}{f^{\gamma}} 
\right)^{2h-2+r},
	\eeq
where $\chi_{\gamma}(p)$ is the character of a $p$-cycle in the
irreducible representation $\gamma$ of $S_p$ and $f^{\gamma}$ is the
dimension of $\gamma$.  For $p=2$, $S_2$ is isomorphic to $\QZ_2$, and
the characters of the transposition for 
two one-dimensional irreducible representations are
$1$ and $-1$, respectively.   It follows that
	\beq
	N_{2,h,r} = T_{2,h,r} =  \left\{ \begin{array}{ll} 
	2^{2h} & \mbox{ for $r$  even\, ,} \\ 
		0 & \mbox{ for $r$ odd\, ,}
	\end{array}\right.  
	\eeq
and therefore that
	\beq
\tilde{\mu}^{g,2}_{h,2}(1,1) = N_{2,h,r} = 2^{2h} \, ,
	\eeq 
which is the desired result.	{\hfill $\qed$}

{\bf Remark:} The answer for the case $g=1$ and $h=1$  is 3, 
which follows from Lemma~\ref{lemma:1}.  For $h=1$ and $g>1$, we have
$\tilde{\mu}^{g,2}_{1,2}(1,1) =4$.

%%%%%%%%%%%%%
%    Degree 3
%%%%%%%%%%%%%

\subsection{Degree Three}
The following lemma will be useful in the ensuing computations:
	\begin{lemma} \label{lemma:b2} Let $t^p_k = 
	2 \, \delta_{k,1}\, \sum_{i=1}^{j} \delta_{p,i}
	+ \delta_{k,2}\, \sum_{i=j+1}^r \delta_{p,i}$.  Then, 
	\beq
	B_{2,h,(t^p_k)} = 	\left\{ \begin{array}{ll} 
	2^{2h} & \mbox{ for $j$  even\, ,} \\ 
		0 & \mbox{ for $j$ odd\, .}
	\end{array}\right.
	\eeq
	\end{lemma}
\noindent
{\sc Proof:}  The result follows trivial 
from the general formula for $B_{n,h,\sigma}$ by
noting that the characters values of the transposition for the
two irreducible representations of $S_2$ are $1$ and $-l$.
{\hfill $\qed$}

We now show
	\begin{claim} The degree-3 simple Hurwitz numbers are given by
	\beq
		N_{3,h,r} = 2^{2h-1} ( 3^{2h-2+r} -1) = 2^{2h-1} (
	3^{2g -4h + 2} -1),
	\eeq
where $r = 6(1-h) + 2(g-1)$ is the number of simple branch points.
	\end{claim}
\noindent
{\sc Proof:}  $T_{3,h,r}$ receives non-zero contributions from the
following 
partitions of 3: $(3)$ and $(1,2)$.  There are three irreducible
representations  
of $S_3$ of dimensions 1,1, and 2, whose respective values of their
characters  
on a transposition are $1,-1,$ and $0$.   Taking care to account for
the 
correct combinatorial factors easily yields the desired result.
{\hfill $\qed$}

%%%%%%%%%%%%
%   Degree 4
%%%%%%%%%%%%

\subsection{Degree Four}
The degree-4 answer is slightly more complicated:
	\begin{claim} The degree-4 simple Hurwitz numbers are given by
	\barray \label{eq:N4}
	N_{4,h,r} &=& 2^{2h-1} \left[ (3^{2h-2 +r}+1) 2^{4h-4+r} - 
	3^{2h-2+r}  - 2^{2h-3+r} +1 \right]  + 2^{4h-4+r}
	(2^{2h}-1)\nonumber\\ 
	          &=& 2^{2h-1} \left[ (3^{2g-6h+4}+1) 2^{ 2g-4h+2} - 
	3^{2g-6h+4}  - 2^{2g-6h+3} +1 \right]  + 2^{2g-4h +2}
	(2^{2h}-1).
\nonumber\\
	\earray
	\end{claim}	

\noindent
{\sc Proof:} The last term in \eqr{eq:N4} comes from the second term in
\eqr{eq:Neven} by applying Lemma~\ref{lemma:Tltwo}.  
To compute $T_{4,h,r}$, we note that the only consistent partitions of 4 and
$\sigma$ are when 4 has the following partitions: $(4), (1,3), (2,2),$ 
and $(1,1,2)$.  The only non-immediate sum involves
	\beq
	\sum_{\sigma_1+ \sigma_2 =\sigma} B_{2,h,\sigma_1} \,
B_{2,h,\sigma_2}  
	\, ,
	\eeq
which, upon applying Lemma~\ref{lemma:b2}, becomes $2^{4h +r -1}$.
{\hfill $\qed$}

\vspace{2mm}

Higher degree computations are similarly executed, 
although one must keep track of some combinatorial factors arising 
from inequivalent distributions of $\sigma$ in \eqr{eq:TB},
and we thus omit their proofs in the subsequent discussions.

%%%%%%%%%%%%%
%    Degree 5
%%%%%%%%%%%%%
\subsection{Degree Five}
For the degree 5 computation, we need
	\begin{lemma}
	Let $t^p_k = 3 \, \delta_{k,1} \,\sum^{j}_{i=1} \delta_{i,p} +
		(\delta_{k,1} + \delta_{k,2})\,\sum^{r}_{i=j+1}$.  Then,
		\beq
		B_{3,h,(t^p_k)} = 	\left\{ \begin{array}{ll} 
			2^{2h}\,3^{2h-1+r-j} & \mbox{ for $j<r$  even\, ,} \\ 
		0 & \mbox{ for $j$ odd\, ,}\\
		2\cdot 3^{2h-1} \, (2^{2h-1}+1) & \mbox{ for $j=r$ \ ,} 
		\end{array}\right.	
		\eeq
	\end{lemma}
from which follows	
	\begin{claim} The degree 5 simple Hurwitz numbers are given by
		\barray
		N_{5,h,r} &=& 2^{2h-1} \,(2^{2h + r-2} - 2^{4h + r-4} -1) 
		-\ 3^{2h-2} \,2^{2h-1} \,( 1  + 2^{2h + r-2}  + 2^{2h
		+ 2r-2}) + \nonumber\\ 
			&&+\ 3^{2h+r-2}\, 2^{2h-1}\, ( 1 - 2^{4h + r-4})  
				+ 2^{6h + r-5}\, 3^{2h-2} + \nonumber\\
			&&	+\ (1 + 2^{4h + r-4} )\, 2^{2h-1}\,
		3^{2h-2} \, 5^{2h + r-2}  .
		\earray
	\end{claim}

%%%%%%%%%%%%%
%    Degree 6
%%%%%%%%%%%%%
\subsection{Degree Six}
Similarly, by using
	\begin{lemma}
	Let $t^p_k = 4 \, \delta_{k,1} \,\sum^{j}_{i=1} \delta_{i,p} +
			(\delta_{k,1} + \delta_{k,2})\,\sum^{r}_{i=j+1}$.  Then,
			\beq
			B_{4,h,(t^p_k)} = 	\left\{ \begin{array}{ll} 
				3\cdot 2^{r-j +6h-2}\,(3^{2h-2+r-j}+1) 
				& \mbox{ for $j<r$  even\, ,} \\ 
			0 & \mbox{ for $j$ odd\, ,}\\
			3\cdot 2^{4h-1} \,(2^{2h-1} 3^{2h-2} + 2^{2h-1} + 3^{2h-2})	
			& \mbox{ for $j=r$ \ ,} 
			\end{array}\right.	
			\eeq
	\end{lemma}
and we obtain	
	\begin{claim}  The degree 6 simple Hurwitz numbers are given by
	\barray	
	 N_{6,h,r} &=&\frac{1}{720} \left[ 360\cdot {2^{2\,h}} - 
	 135\cdot {2^{4\,h + r}} -
       40\cdot{2^{2\,h}}\cdot{3^{2\,h + r}} -
       {{5\cdot{2^{2\,h}}\cdot{3^{4\,h + r}}\,
           \left( 8 + {2^{2\,h + r}} \right) }\over 9} \right. \nonumber \\
		   &&+
       20\cdot{2^{2\,h}}\cdot{3^{2\,h}}\,
        \left( 4 + {2^{2\,\left( h + r \right) }} + {2^{2\,h + r}} \right)  +
       {{15\cdot{2^{6\,h}}\,\left( 3 + {3^r} \right) }\over 2} \nonumber\\
	   &&+
       {{5\cdot{2^{6\,h + r}}\,\left( 9 + {3^{2\,h + r}} \right) }\over 2}-
       {{{2^{2\,h}}\cdot{3^{2\,h}}\,
           \left( 25\cdot{2^{4\,h + r}} + 16\cdot{5^{2\,h + r}} +
             {2^{4\,h + r}}\cdot{5^{2\,h + r}} \right) }\over {10}} 
			 \nonumber\\
			 &&+
       {{{2^{6\,h}}\,\left( 100\cdot{3^{4\,h + r}} +
             25\cdot{2^{2\,h}}{3^{4\,h + r}} +
             25\cdot{2^{2\,h}}{3^{4\,h + 2\,r}} +
             81\cdot{2^{2\,h}}{5^{2\,h + r}} +
             {2^{2\,h}}{3^{4\,h + r}}{5^{2\,h + r}} 
			 \right) }\over {360}}
			 \nonumber\\
        &&-\left. {{5\cdot{2^{6\,h}}\left( 9\cdot{2^{2\,h}} + 4\cdot{3^{2\,h}} +
             9\cdot{2^{2\,h}}{3^r} + {2^{2\,h}}{3^{2\,h}}{5^r} +
             {2^{2\,h}}{3^{2\,h}}{7^r} \right) }\over 8} 
			 \right]\nonumber\\
		&& +\, 2^{6h-5} 3^{r-1} 
		\left[ 3^{2h-1} (2^{2h-1} +1) -3 (2^{2h-1}) +1\right].
		\earray
		\end{claim}

%%%%%%%%%%%%%
%    Degree 7
%%%%%%%%%%%%%
\subsection{Degree Seven}

	\begin{claim} The degree 7 simple Hurwitz numbers are given by
	\barray
	N_{7,h,r} &=&
    {{-{2^{2\,h}}}\over 2} - {{3\,{2^{6\,h}}}\over {32}} +
    {{{2^{8\,h}}}\over {64}} + {{{2^{4\,h + r}}}\over 4} -
    {{{2^{6\,h + r}}}\over {32}} - {{{2^{2\,h}}\,{3^{2\,h}}}\over 6} +
    {{{2^{6\,h}}\,{3^{2\,h}}}\over {96}} \nonumber \\
	& &\ -
    {{{2^{8\,h}}\,{3^{2\,h}}}\over {576}} -
    {{{2^{4\,h + r}}\,{3^{2\,h}}}\over {36}} -
    {{{2^{4\,h + 2\,r}}\,{3^{2\,h}}}\over {24}} +
    {{{2^{6\,h + r}}\,{{{5\over 2}}^r}\,{3^{2\,h}}}\over {288}} -
    {{{2^{8\,h + r}}\,{{{5\over 2}}^r}\,{3^{2\,h}}}\over {1152}} -
    {{{2^{6\,h}}\,{3^r}}\over {32}} \nonumber\\
	&& \ + {{{2^{8\,h}}\,{3^r}}\over {64}} +
    {{{2^{2\,h}}\,{3^{2\,h + r}}}\over {18}} +
    {{{2^{6\,h}}\,{3^{2\,h + r}}}\over {144}} -
    {{{2^{8\,h}}\,{3^{2\,h + r}}}\over {1152}} -
    {{{2^{6\,h + r}}\,{3^{2\,h + r}}}\over {288}} \nonumber \\
	&&\ +
    {{{2^{2\,h}}\,{3^{4\,h + r}}}\over {81}} -
    {{{2^{6\,h}}\,{3^{4\,h + r}}}\over {1296}} -
    {{{2^{8\,h}}\,{3^{4\,h + r}}}\over {10368}} +
    {{{2^{4\,h + r}}\,{3^{4\,h + r}}}\over {648}} -
    {{{2^{6\,h + r}}\,{3^{4\,h + r}}}\over {2592}} \nonumber\\
	& &\ -
    {{{2^{8\,h}}\,{3^{4\,h + 2\,r}}}\over {5184}} -
    {{{2^{6\,h}}\,{5^{2\,h}}}\over {800}} +
    {{{2^{8\,h}}\,{3^{2\,h}}\,{5^{2\,h}}}\over {28800}} -
    {{{2^{4\,h + 2\,r}}\,{3^{2\,h}}\,{5^{2\,h}}}\over {1800}} -
    {{{2^{8\,h}}\,{3^{2\,\left( h + r \right) }}\,{5^{2\,h}}}\over {28800}}\nonumber\\
	&&\ -
    {{{2^{4\,h + r}}\,{3^{2\,h + r}}\,{5^{2\,h}}}\over {1800}} +
    {{{2^{6\,h + r}}\,{3^{4\,h + r}}\,{5^{2\,h}}}\over {64800}} +
    {{{2^{6\,h}}\,{3^{4\,h + 2\,r}}\,{5^{2\,h}}}\over {64800}} +
    {{{2^{8\,h}}\,{3^{2\,h}}\,{5^r}}\over {576}} \nonumber\\
	&& \ -
    {{{2^{8\,h}}\,{5^{2\,h + r}}}\over {3200}}  +
    {{{2^{2\,h}}\,{3^{2\,h}}\,{5^{2\,h + r}}}\over {450}} +
    {{{2^{6\,h + r}}\,{3^{2\,h}}\,{5^{2\,h + r}}}\over {7200}} -
    {{{2^{8\,h}}\,{3^{4\,h + r}}\,{5^{2\,h + r}}}\over {259200}}\nonumber\\ 
	&& \ +
    {{{2^{8\,h}}\,{3^{2\,h}}\,{7^r}}\over {576}}  +
    {{{2^{8\,h}}\,{3^{2\,h}}\,{7^{2\,h + r}}}\over {56448}} +
    {{{2^{6\,h + r}}\,{3^{2\,h}}\,{5^{2\,h}}\,{7^{2\,h + r}}}\over
      {352800}} + {{{2^{8\,h}}\,{3^{4\,h + r}}\,{5^{2\,h}}\,
        {7^{2\,h + r}}}\over {12700800}} \nonumber\\
		&& \ -
    {{{2^{8\,h}}\,{3^{2\,h}}\,{5^{2\,h}}\,{{11}^r}}\over {28800}}.
	\earray
	\end{claim}

%%%%%%%%%%%%%%%%
%
%%%%%%%%%%%%%%%%

\setall
\section{Reducible Covers}\label{app:red}

\barray
	B_{n,h,r} &=& (n!)^{2h-1} \left(\hspace{-1mm}\begin{array}{l}
		n\\2 \end{array}\hspace{-1mm}\right)^r \, \left[	
		\sum_{\gamma\in {\SR_n}} \frac{1}{(f^{\gamma})^{2h-2}} \left(
		\frac{\,\chi_{\gamma}(2)}{f^{\gamma}}\right)^r
		\right]\, , \nonumber\\
	B_{2,h,r} &=& 2\cdot 2^{2h-1}\, , \nonumber\\
	B_{3,h,r} &=& 2\cdot 3!^{2h-1} \left(\hspace{-1mm}\begin{array}{l}
		3\\2 \end{array}\hspace{-1mm}\right)^r  \, , \nonumber\\
	B_{4,h,r} &=& 2\cdot 4!^{2h-1} \left(\hspace{-1mm}\begin{array}{l}
		4\\2 \end{array}\hspace{-1mm}\right)^r \left[  1+
		\frac{1}{3^{2h-2+r}}
		\right]  \, , \nonumber\\
	B_{5,h,r} &=& 2\cdot 5!^{2h-1} \left(\hspace{-1mm}\begin{array}{l}
		5\\2 \end{array}\hspace{-1mm}\right)^r \left[  1+
		\frac{2^r}{4^{2h-2+r}} +\frac{1}{5^{2h-2+r}} 
		\right]  \, , \nonumber\\
	B_{6,h,r} &=& 2\cdot 6!^{2h-1} \left(\hspace{-1mm}\begin{array}{l}
		6\\2 \end{array}\hspace{-1mm}\right)^r \left[  1+
		\frac{3^r}{5^{2h-2+r}} +\frac{3^r}{9^{2h-2+r}} + 
		 \frac{2^r}{10^{2h-2+r}} +\frac{1}{5^{2h-2+r}}
		\right]  \, , \nonumber\\
	B_{7,h,r} &=& 2\cdot 7!^{2h-1} \left(\hspace{-1mm}\begin{array}{l}
		7\\2 \end{array}\hspace{-1mm}\right)^r \left[  1+
		\frac{4^r}{6^{2h-2+r}} +\frac{6^r}{14^{2h-2+r}} + 
		 \frac{5^r}{15^{2h-2+r}}
		+\frac{4^r}{14^{2h-2+r}}\right.
		\nonumber\\
		&&\left. \ + \frac{5^r}{35^{2h-2+r}} +\frac{1}{21^{2h-2+r}}
		\right]  \, , \nonumber\\
	B_{8,h,r} &=& 2\cdot 8!^{2h-1} \left(\hspace{-1mm}\begin{array}{l}
		8\\2 \end{array}\hspace{-1mm}\right)^r \left[  1+
		\frac{5^r}{7^{2h-2+r}} +\frac{10^r}{20^{2h-2+r}} + 
		 \frac{9^r}{21^{2h-2+r}}
		+\frac{10^r}{28^{2h-2+r}}\right.
		\nonumber\\
		&&\left. \ + \frac{16^r}{64^{2h-2+r}}
		+\frac{5^r}{35^{2h-2+r}}
		+ \frac{4^r}{14^{2h-2+r}} +\frac{10^r}{70^{2h-2+r}}
		+\frac{4^r}{56^{2h-2+r}}
		\right]  \, , \nonumber\\
	B_{9,h,r} &=& 2\cdot 9!^{2h-1} \left(\hspace{-1mm}\begin{array}{l}
		9\\2 \end{array}\hspace{-1mm}\right)^r \left[  1+
		\frac{6^r}{8^{2h-2+r}} +\frac{15^r}{27^{2h-2+r}} + 
		 \frac{14^r}{28^{2h-2+r}} +\frac{20^r}{48^{2h-2+r}}\right.
		\nonumber\\
		&& \ 
		+ \frac{35^r}{105^{2h-2+r}} +\frac{14^r}{56^{2h-2+r}}
		+ \frac{14^r}{42^{2h-2+r}} +\frac{36^r}{162^{2h-2+r}}
		+\frac{20^r}{120^{2h-2+r}}\nonumber\\
		&&\left.
		\ + \frac{21^r}{189^{2h-2+r}} +\frac{14^r}{84^{2h-2+r}}
		+ \frac{14^r}{168^{2h-2+r}} +\frac{6^r}{216^{2h-2+r}}
		\right]  \, , \nonumber\\
	B_{10,h,r} &=& 2\cdot 10!^{2h-1} \left(\hspace{-1mm}\begin{array}{c}
		10\\2 \end{array}\hspace{-1mm}\right)^r \left[  1+
		\frac{7^r}{9^{2h-2+r}} +\frac{21^r}{35^{2h-2+r}} + 
		 \frac{20^r}{36^{2h-2+r}} +\frac{35^r}{75^{2h-2+r}}\right.
		\nonumber\\
		&& \ 
		+ \frac{64^r}{160^{2h-2+r}} +\frac{28^r}{84^{2h-2+r}}
		+ \frac{34^r}{90^{2h-2+r}} +\frac{91^r}{315^{2h-2+r}}
		+\frac{55^r}{225^{2h-2+r}}\nonumber\\
		&& \ 
		+ \frac{70^r}{350^{2h-2+r}} +\frac{14^r}{126^{2h-2+r}}
		+ \frac{14^r}{42^{2h-2+r}} +\frac{64^r}{288^{2h-2+r}}
		+\frac{70^r}{450^{2h-2+r}}\nonumber\\
		&&\left.
		\ + \frac{63^r}{567^{2h-2+r}} +\frac{35^r}{525^{2h-2+r}}
		+ \frac{28^r}{252^{2h-2+r}} +\frac{20^r}{300^{2h-2+r}}
		+\frac{14^r}{210^{2h-2+r}}
		\right]  \, . \nonumber\\
\earray

%%%%%%%%%%%%%%%
%
%%%%%%%%%%%%%%%
\setall
\section{Simple Hurwitz Numbers for an Elliptic Curve Target}  \label{app:elliptic curve}
We can compare our answers in the case of an elliptic curve target
with those obtained from string theory.  To do so,
we organize $T_{n,1,2g-2}/n!$ into a generating function $H^g_1(q)$,
which is defined as
\[ (2g-2)! H^g_1 = \sum^\infty_{n=1} \mu^{g,n}_{1,n}\, q^n =
\sum^\infty_{n=1}\frac{T_{n,1,2g-2}}{n!} q^n . \]
Our explicit formulas for $T_{n,1,2g-2}/n!$, $n\leq 7$, from \S\ref{app:comp} and the
recursive method discussed in \S\ref{sec: recursion} give rise to the
following $q$-expansions of $H^g_1(q)$:

{\small
        \barray
  2!\, H^2_1 &=&  2 q^2 + 16 q^3 + 60 q^4 + 160 q^5 + 360 q^6 + 672
q^7 + 1240 q^8 + 1920 q^9 + 3180 q^{10} + {\cal O}(q^{11}) \, , \nonumber\\
%%%%%%%%%%
4! \, H^3_1 &=& 2 q^2 + 160 q^3 + 2448 q^4 + 18304 q^5 + 90552 q^6 +
\nonumber\\
& &+ 341568 q^7 + 1068928 q^8 + 2877696 q^9 + 7014204 q^{10} +
{\cal O}(q^{11}) \, , \nonumber\\
%%%%%%%%%%
6!\, H^4_1 &=& 2 q^2 + 1456 q^3 + 91920 q^4 + 1931200 q^5 +
21639720 q^6 +  \nonumber\\
&&+160786272 q^7 + 893985280 q^8 + 4001984640 q^9 + 15166797900 q^{10} +
{\cal O}(q^{11})\, , \nonumber\\
%%%%%%%%%%
8!\, H^5_1 &= &2 q^2 + 13120 q^3 + 3346368 q^4 + 197304064 q^5 +
 5001497112 q^6 +
 \nonumber\\
&&+ 73102904448 q^7 + 724280109568 q^8 + 5371101006336 q^9 + \nonumber\\
&&+ 31830391591644 q^{10} + {\cal O}(q^{11}) \, , \nonumber\\
%%%%%%%%%%
10!\, H^6_1 &= & 2 q^2 + 118096 q^3 + 120815280 q^4 +
19896619840 q^5 + 1139754451080 q^6 + \nonumber\\
&& + 32740753325472 q^7 + 577763760958720 q^8 + 7092667383039360 q^9
+ \nonumber\\
&& + 65742150901548780 q^{10} + {\cal O}(q^{11}) \, , \nonumber\\
%%%%%%%%%%
12!\,
H^7_1 &=&  2q^2 +  1062880 q^3 + 4352505888 q^4+  1996102225024 q^5+
258031607185272 q^6 + \nonumber\\
&& +
14560223135464128 q^7+ 457472951327051008 q^8+ 9293626316677061376 q^9
+ \nonumber\\
&& + 134707212077147740284 q^{10} + {\cal O}(q^{11}) \, , \nonumber\\
%%%%%%%%%%
14!\, H^8_1 &= &2 q^2 + 9565936 q^3 + 156718778640 q^4 + 199854951398080 q^5 +
58230316414059240 q^6  + \nonumber\\
&& + 6451030954702152672 q^7 + 360793945093731688960q^8 + \nonumber\\
&& + 12127449147074861834880q^9 + 274847057797905019237260 q^{10} + {\cal
O}(q^{11})\, , \nonumber\\
%%%%%%%%%%
16!\, H^9_1 &= & 2q^2  + 86093440q^3  +5642133787008q^4
+19994654452125184q^5 +  \nonumber\\
&&+13120458818999011032q^6
+2852277353239208548608q^7 + \nonumber\\
&& +283889181859169785013248q^8
+15786934495235533394850816q^9 + \nonumber\\ &&
+559374323532926110389380124q^{10}+
{\cal O}(q^{11})\, , \nonumber\\
%%%%%%%%%%
18!\, H^{10}_1 &=&   2q^2  + 774840976q^3  + 203119138758000q^4  +
1999804372817081920q^5 + \nonumber\\
&&+ 2954080786719122704200q^6
+1259649848110685616355872q^7 + \nonumber\\
&&+223062465532295875789024000q^8
+20519169517386068841434851200q^9 + \nonumber\\
&&+1136630591006374329359969015340
q^{10} + {\cal O}(q^{11})\, , \nonumber\\
%%%%%%%%%%
20! H^{11}_1& = & 2q^2  + 6973568800q^3  +7312309907605728q^4
+199992876225933468544q^5 +  \nonumber\\
&&+664875505232132669710392q^6
+555937950399900003838125888q^7 + \nonumber\\
&& +175116375615275397674821996288q^8
+26643243663812779066608784102656q^9 + \nonumber\\
&&+2307123097757961461530407199135164 q^{10} + {\cal
O}(q^{11})\, , \nonumber\\
%%%%%%%%%%
22! H^{12}_1 &=& 2q^2  +62762119216q^3  +263243344926609360q^4
+19999741489842287527360q^5  + \nonumber\\
&&+149618514702670218774465960q^6
+245271669454107089851705983072q^7 + \nonumber\\
&&+137402588289598470102013264291840q^8  + \nonumber\\
&&+34572266592868474818152471335048320q^9 +
\nonumber\\ &&+4679534045992767568052180827613155020 q^{10} + {\cal
O}(q^{11})\, . \nonumber
        \earray }

The free energies  $\SF_g$
of string theory on the target space of an elliptic
curve are known to be quasi-modular forms of weight $6g-6$.
They have been computed up to genus 8 in \cite{Rudd} and have the same
expansions in $q=\exp (\hat{t})$, where $\hat{t}$ is the K\"{a}hler
parameter of the elliptic curve, as what we have above for $H^g_1$.

For convenience, we also summarize the simple Hurwitz numbers for an
elliptic curve target and arbitrary source Riemann surfaces 
up to degree 7:
\barray
	\mu^{g,1}_{1,1} (1) &=& \delta_{g,1}\, , \nonumber\\
	 \mu^{g,2}_{1,2} (1^2) &=& 2\, , \nonumber\\
	\mu^{g,3}_{1,3} (1^3) &=& 2\, \left[ 3^r-1 \right]\, ,  \nonumber\\
	\mu^{g,4}_{1,4} (1^4) &=&  2\,\left[ 6^r +  2^{r-1}
	- 3^r +1   \right]\, ,
	  \nonumber\\
	\mu^{g,5}_{1,5} (1^5) &=& 2 \, \left[ 10^r  -6^r + 5^r -4^r+
	3^r-2 \right]\, ,
	\nonumber\\
	 \mu^{g,6}_{1,6} (1^6) &=& 2\cdot {{15}^r} - 2\cdot {{10}^r} +
	2\cdot {9^r} -  2\cdot {7^r} + {6^r} - 2\cdot {5^r}+ 4\cdot 
	 {4^r}  -4 \cdot 3^{r-1} + {2^r} +4 \, ,    \nonumber\\
	\mu^{g,7}_{1,7} (1^7) &=& 2\,\left[  21^r -15^r +14^r-11^r
	+10^r -2 \cdot 9^r +3\cdot 7^r- {6^r} + 2\cdot 5^r
	-4\cdot{4^{r}}\right. + \nonumber\\
	&& \left. \ +\,
	2\cdot 3^r - {2^r} -4
	  \right] , \label{eq:e1}
\earray
where $r=2g-2$.

%%%%%%%%%%%%%%%%
%
%%%%%%%%%%%%%%%%
\section{The Hurwitz Numbers $\mu^{g,2k}_{0,2} (k,k)$} \label{sec:app mu2k}
	
\barray
	\mu^{1,2k}_{0,2} (k,k) &=& {(2k+2)!\over 48 \,(k!)^2} 
	           \,(3k-2)\, k^{2k+1} \, , \nonumber\\
	\mu^{2,2k}_{0,2} (k,k) &=& { (2k+4)! \over 2\, (k!)^2} \, 
	\left({ 28 - 73 \, k + 49 k^2 \over 2880}\right)\, k^{2k+3} 
		\, , \nonumber\\
	\mu^{3,2k}_{0,2} (k,k) &=& { (2k+6)! \over 2\, (k!)^2} \, 
	\left(  {{-744 + 2530\,k - 2949\,{k^2} + 1181\,{k^3}}
	\over {725760}}\right)\, k^{2k+5} \, , \nonumber\\
	\mu^{4,2k}_{0,2} (k,k) &=& { (2k+8)! \over 2\, (k!)^2} \, 
	    	\left(  {{18288 - 72826\,k + 111309\,{k^2} - 77738\,{k^3}
	    	+ 21015\,{k^4}}\over   {174182400}}
            \right)\, k^{2k+7} \, , \nonumber\\
	\mu^{5,2k}_{0,2} (k,k) &=& { (2k+10)! \over 2\, (k!)^2} \,
		k^{2k+9} {1 \over 22992076800} \left(		
			    	 -245280 + 1086652\,k -
		1959376\,{k^2}  \right. \nonumber \\
&& \hspace{1cm} \left. + 1807449\,{k^3} - 857552\,{k^4} +
		168155\,{k^5} \right) \, , \nonumber\\ 
	\mu^{6,2k}_{0,2} (k,k) &=& { (2k+12)! \over 2\, (k!)^2} \, k^{2k+11} 
	\, {1\over  {753220435968000}}\, \left(   
		814738752 - 3904894152\,k\right.\nonumber\\
		&&\hspace{1cm} +\, 7889383898\,{k^2} - 8650981635\,{k^3} +
      5462073347\,{k^4} \nonumber\\
&&\hspace{1cm} \left. - 1892825445\,{k^5} + 282513875\,{k^6}
\right) \, . \nonumber\\
\earray

%%%%%%%%%%%%%%%%%%
%  REFERENCES
%%%%%%%%%%%%%%%%%%

\end{document}